\documentclass[11pt, a4paper, twocolumn]{article}
\pdfoutput=1
\usepackage{a4wide}
\usepackage[utf8]{inputenc}
\usepackage[T1]{fontenc}
\usepackage{mlmodern}
\usepackage{textcomp}
\usepackage[english]{babel}
\usepackage{microtype}
\usepackage{csquotes}

\usepackage{graphicx}
\usepackage{subcaption}
\usepackage{cancel}

\usepackage[dvipsnames, svgnames, table]{xcolor}
\selectcolormodel{cmy}

\usepackage[
	backend=bibtex8,
	citestyle=numeric-comp,
	sorting=none, sortcase=false, sortcites=true,
	giveninits=true,
	minnames=10,
	maxnames=10,
	date=year
]{biblatex}

\AtEveryBibitem{
  \clearfield{issn}
  \clearfield{language}
  \ifentrytype{article}{
    \clearfield{publisher}
    \clearfield{note}
  }{}
}
\DeclareFieldFormat[article,inbook,incollection,inproceedings,book,thesis,report,misc]
  {title}{\mkbibemph{#1}}

\usepackage{amsmath, amsfonts, amssymb}
\usepackage{braket}
\usepackage{siunitx}

\usepackage[
	breaklinks=true, linktocpage, colorlinks=true,
	urlcolor=RoyalBlue, linkcolor=RoyalBlue, citecolor=BrickRed
]{hyperref}
\usepackage{hyperxmp}
\usepackage[all]{hypcap}
\usepackage{xurl}

\usepackage[heightrounded,
    a4paper,
	top=2cm, bottom=2cm, left=2.5cm, right=2.5cm,
	marginparwidth=1.75cm
]{geometry}
 
\usepackage{authblk}

\newcommand{\email}[1]{\href{mailto:#1}{\nolinkurl{#1}}}
\newcommand{\emailfoot}[1]{\thanks{\email{#1}}}

\newcommand{\rootpath}{./}
\graphicspath{{\rootpath}}
\numberwithin{equation}{section}

\usepackage[sort&compress, english]{cleveref}
\crefname{figure}{Fig.}{Figs.}     \Crefname{figure}{Figure}{Figures} \crefname{equation}{Eq.}{Eqs.}     \Crefname{equation}{Equation}{Equations} \crefname{section}{Sec.}{Secs.}     \Crefname{section}{Section}{Sections} \Crefname{appsec}{Appendix}{Appendices}\crefname{appsec}{App.}{App.}

\input{arxiv.def}

\hypersetup{
	pdftitle={Many-Body Quantum Score: a scalable benchmark for digital and analog quantum processors and first test on a commercial neutral atom device},
	pdflang={en},
}

\title{Many-Body Quantum Score: a scalable benchmark for digital and analog quantum processors and first test on a commercial neutral atom device}

\author[1]{Harold Erbin\emailfoot{harold.erbin@cea.fr}}
\author[2]{Pierre-Louis Burdeau\emailfoot{pl.burdeau@gmail.com}}
\author[2]{Corentin Bertrand\emailfoot{corentin.bertrand@eviden.com}}
\author[3,2]{Thomas Ayral\emailfoot{thomas.ayral@polytechnique.edu}}
\author[1]{Grégoire Misguich\emailfoot{gregoire.misguich@cea.fr}\footnote{G. Misguich is the corresponding author.}}

\affil[1]{Université Paris-Saclay, CEA, CNRS, Institut de Physique Théorique
	\protect\\
	91191 Gif-sur-Yvette, France
}
\affil[2]{Eviden Quantum Lab, 78340 Les Clayes-sous-Bois, France
}
\affil[3]{CPHT, CNRS, École Polytechnique, IP Paris, 91128 Palaiseau, France
}

\newcommand{\rev}[1]{#1} 

\renewcommand{\vec}[1]{\boldsymbol{#1}}
\newcommand{\e}[0]{\mathrm{e}}
\newcommand{\I}[0]{\mathrm{i}}
\newcommand{\dd}[0]{\mathrm{d}}

\date{September 10, 2026}

\begin{document}

\maketitle

\begin{abstract}
\rev{We propose the Many-Body Quantum Score (MBQS), a practical and scalable application-level benchmark protocol designed to evaluate how accurately quantum processing units (QPUs)---both gate-based and analog---can simulate many-body quantum dynamics. MBQS quantifies performance by identifying the maximum number of qubits with which a QPU can reliably reproduce correlation functions of the transverse-field Ising model following a specific quantum quench. This quench reveals a global correlation surge, a collective peak of connected correlations at all distances, which to our knowledge had not been noticed so far in this context. This paper presents the MBQS protocol and highlights its design principles, supported by analytical insights, classical simulations, and experimental data. It also reports an implementation on Ruby, an analog QPU based on Rydberg atoms developed by the Pasqal company. These findings demonstrate MBQS's potential as a robust and informative tool for benchmarking near-term quantum devices for many-body physics. While the present experimental validation has been performed on an analog device, the protocol has the potential to be implemented on digital platforms as well.}
\end{abstract}

\newpage

\tableofcontents

\section{Introduction}

Benchmarking has been an important driver of technological progress in many fields, notably machine learning~\cite{Dai:2019:BenchmarkingContemporaryDeep,Reddi:2020:MLPerfInferenceBenchmark,Thiyagalingam:2022:ScientificMachineLearning} and high-performance computing (HPC)~\cite{Ihde:2021:SurveyBigData}.
For example, supercomputers are compared in the TOP500 ranking, and the ILSVRC challenge prompted breakthroughs in deep learning in 2012.
Benchmarks provide a (relatively) objective comparison between different systems, making it possible to measure capabilities and performance.
From a user's perspective, benchmarks help users make informed decisions by providing clear reference points for matching platforms to tasks, i.e.~selecting the most appropriate system to solve a problem while taking into account its strengths and weaknesses.
From a provider's perspective, benchmarks provide targets for product development and can guide R\&D by identifying bottlenecks and failing components, setting performance goals, and validating upgrades.
Publishing benchmark results increases trust and transparency between users and providers while also ensuring reproducibility.

\begin{figure*}
    \centering
    \fbox{\includegraphics[width=\linewidth]{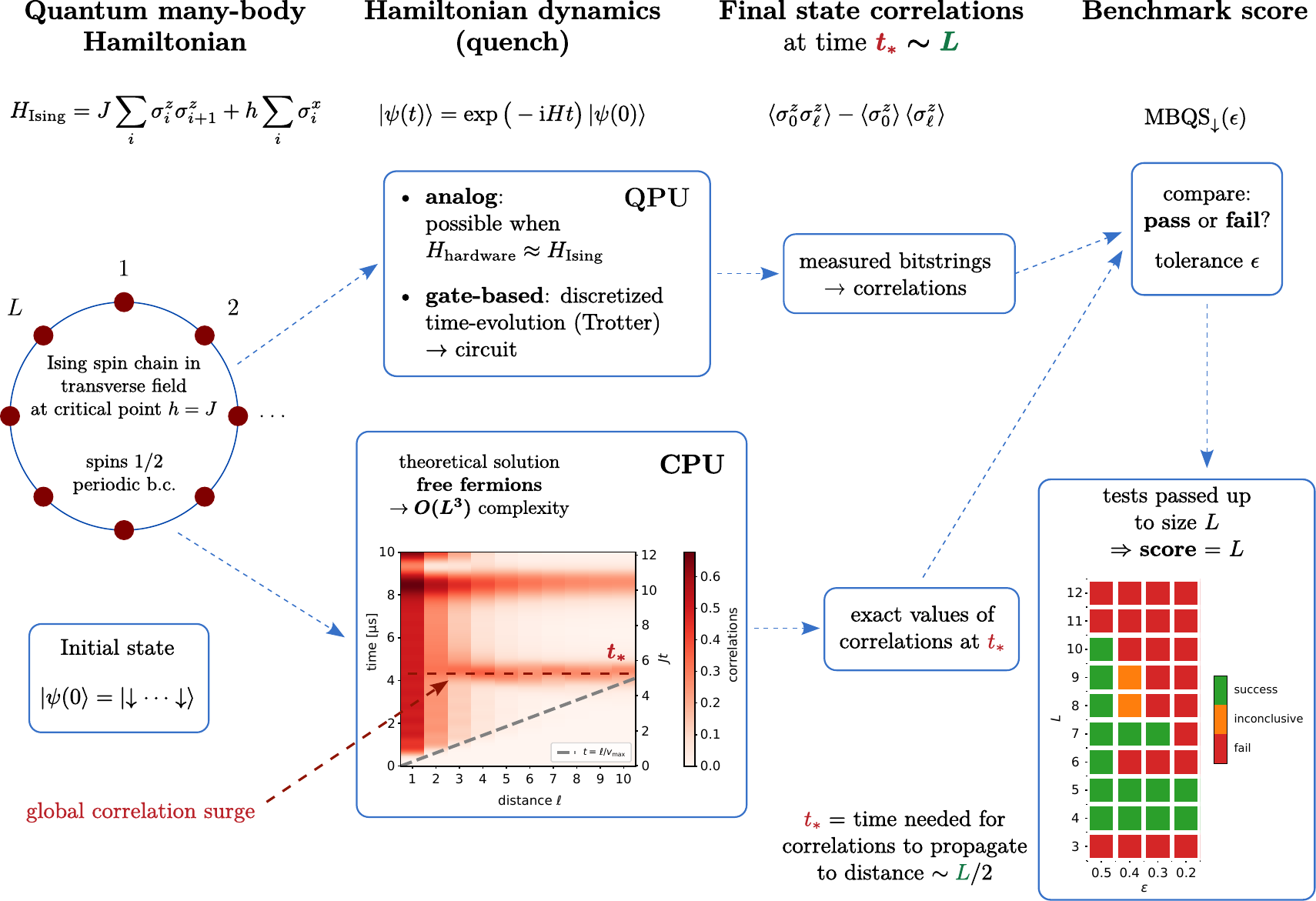}}
    \caption{Proposed benchmark protocol.
    }
    \label{fig:summary}
\end{figure*}

In quantum computing, scalability of benchmarks is an essential requirement.
Even though present (noisy) quantum devices can still often be emulated by classical computers running state-of-the-art algorithms, some technologies provide devices which can operate close to the limit of classical simulability~\cite{bernien_probing_2017,lienhard_observing_2018,ebadiQuantumPhasesMatter2021,KIM_EvidenceUtilityQuantum_2023,HAGHSHENAS_DigitalQuantumMagnetism_2025,NIGMATULLIN_ExperimentalDemonstrationBreakeven_2025}.
We need benchmarks that can quantify the performance of current and \emph{future} quantum processors.
It is therefore important to develop scalable benchmarking protocols, meaning that their resource requirements should not grow exponentially with system size.

Our objective is to propose a scalable benchmark in the field of quantum physics simulations, for both analog and digital quantum processing units (QPUs).
Indeed, the simulation of many-body quantum Hamiltonians is one of the fields where quantum computers are expected to bring some exponential advantage over classical ones~\cite{feynmanSimulatingPhysicsComputers1982,lloydUniversalQuantumSimulators1996,KASSAL_SimulatingChemistryUsing_2011,GEORGESCU_QuantumSimulation_2014,BAUER_HybridQuantumClassicalApproach_2016,preskillQuantumComputingNISQ2018,BAUER_QuantumAlgorithmsQuantum_2020,ayral_quantum_2023}.

To date, most benchmarking efforts have focused on digital devices for an obvious reason: since they are intended to be
universal, one expects (or hopes) them to become the dominant type of device in the future.
Moreover, their computations consist of sequential discrete steps that can be described using group theory and binary logic and are therefore readily amenable to characterization and abstraction.
On the other hand, analog devices are in general not universal,\footnote{
	Note that some works have studied how such platforms could also be used as universal quantum computers~\cite{Cubitt:2018:UniversalQuantumHamiltonians,Kohler:2022:GeneralConditionsUniversality,CESA_UniversalQuantumComputation_2023,Hu:2025:UniversalDynamicsGlobally}.
}
which implies that they offer less abstraction, are more technology-dependent, and generally consist of a single continuous-time evolution process.

However, one can expect analog devices to be more efficient in the short to medium term for solving certain categories of problems, such as graph-based and combinatorial problems and many-body problems, since they directly encode many-body Hamiltonians with spin, bosonic, or fermionic degrees of freedom~\cite{Ayral2025}.
For example, analog QPUs currently offer hundreds to thousands of qubits through neutral atom platforms~\cite{henrietQuantumComputingNeutral2020,ebadiQuantumPhasesMatter2021,Pichard:2024:RearrangementSingleAtoms, Chiu:2025:ContinuousOperationCoherent, Manetsch:2025:TweezerArray6100} or superconducting annealers~\cite{McGeoch:2022:DWaveAdvantage2Prototype, DWave:2025:PerformanceGainsD-Wave}.
Reviews on benchmarking quantum computers include~\cite{Eisert:2020:QuantumCertificationBenchmarking, Proctor:2024:BenchmarkingQuantumComputers, Lorenz:2025:SystematicBenchmarkingQuantum, Lall:2025:ReviewCollectionMetrics, Rohe:2025:QuantumComputerBenchmarking}.

\rev{Because analog QPUs are more constrained than gate-based QPUs, we take the analog setting as the starting point for designing a benchmark applicable to both architectures.
A benchmark designed for analog Hamiltonian dynamics can then be implemented on gate-based machines by discretizing or encoding the same target evolution, for instance with Trotterization or block encoding~\cite{Low:2017:OptimalHamiltonianSimulation, Gilyen:2019:QuantumSingularValue}.}
As indicated above, an application benchmark based on many-body physics is also the most suitable in the short-term given its relevance for quantum chemistry, materials science or even nuclear physics.
\rev{To our knowledge, the protocol defined in this paper, the Many-Body Quantum Score (MBQS), is among the first application-level benchmarks intended to be usable on both analog and digital platforms while probing genuine Hamiltonian quantum dynamics.}
\rev{The experiments described below provide a first test of MBQS on a commercial neutral atom QPU.}

One can distinguish at least two main classes of quantum many-body problems: ground-state searches and Hamiltonian dynamics. We focus here on Hamiltonian dynamics, for which the potential advantage of QPUs is clearer than for ground-state problems~\cite{ayral_quantum_2023}.
The benchmark protocol we propose, summarized in \cref{fig:summary}, is based on a global quantum quench in the one-dimensional ($1d$) transverse-field Ising model, a paradigmatic example of an out-of-equilibrium quantum many-body problem.
The initial state is a simple product state. During the Hamiltonian evolution, correlations spread ballistically and quantum entanglement grows. The evolution is stopped at the surge time, a special time proportional to the system size. At that time, the 2-point connected correlations peak regardless of the distance between the two spins. Although protocols of this type have already been studied extensively, this phenomenon, which we call the
\emph{global correlation surge}, seems to have escaped notice so far.
The main idea is to exploit this dynamically generated correlated state to test the capacity of the device to (1) prepare a simple initial state, (2) perform a time evolution with a controlled Hamiltonian, and (3) provide accurate measurements in an entangled and correlated state. In order to define a \emph{scalable} benchmark score, one exploits the exactly solvable nature of the quantum Ising chain (free fermion integrability~\cite{katsuraStatisticalMechanicsAnisotropic1962,mccoyStatisticalMechanicsXY1971,Mbeng_2024}) to obtain reference results in polynomial time on a classical computer. \rev{MBQS is therefore designed as a verifiable benchmark of accuracy, noise sensitivity, Hamiltonian control, coherence, and measurement quality. It is not, by itself, a protocol to demonstrate quantum advantage: on the contrary, classical tractability of the reference values is required for an unambiguous benchmark.} This benchmark is applicable to current noisy intermediate-scale quantum (NISQ) devices, and we present an implementation on a commercial analog neutral atom device from Pasqal. Thanks to the scalability of the benchmark, it will also be possible to test the performance of future machines with error correction on larger instances of the problem, which require more qubits, greater circuit depth, and higher precision.

\paragraph{Other benchmarks for Hamiltonian simulations}

The benchmark protocol proposed by Granet \emph{et al.}~\cite{GRANET_AppQSimApplicationorientedBenchmarks_2025} shares some similarities with ours: it focuses on free fermion Hamiltonian dynamics, and the final evolution time is proportional to the system size. However, its Hamiltonian is formulated in terms of fermionic rather than spin operators and therefore cannot be simulated directly on an analog neutral atom platform. Another important difference is that the model is defined on a two-dimensional ($2d$) lattice.
Finally, Ref.~\cite{GRANET_AppQSimApplicationorientedBenchmarks_2025} focuses on a local observable, whereas we test the ability of the machine to reproduce correlations at all distances, which are relevant physical observables for distinguishing states of matter. We also stress that long-distance correlations are more difficult to estimate using classical methods than short-distance quantities, particularly when relying on a variational method.

A recent preprint~\cite{aguirreQuSquareScalableQualityOriented2025} proposes a benchmark based on the Hamiltonian simulation of the exactly solvable quantum Ising chain, as does our protocol. However, as in the work mentioned above, the authors focus on a \emph{local} observable (the transverse magnetization) rather than on correlations. Another important difference is that the evolution time is not known a priori and must be determined through trial and error. Finally, the protocol of Ref.~\cite{aguirreQuSquareScalableQualityOriented2025} cannot be implemented directly on current Rydberg simulators because their native measurement basis is longitudinal (along the axis of the Ising interaction) rather than aligned with the transverse field.

Reference~\cite{MCCASKEY_QuantumChemistryBenchmark_2019} proposed a benchmark based on variationally estimating the ground-state energy of an electronic-structure problem in chemistry.
Reference~\cite{DALLAIRE-DEMERS_ApplicationBenchmarkFermionic_2020} also considered a ground-state-energy problem addressed with a variational algorithm. The Hamiltonian is the $1d$ Fermi--Hubbard model, which has an exact solution via the Bethe ansatz.
We also mention Ref.~\cite{dongQuantumHamiltonianSimulation2022}, which introduced a benchmark for Hamiltonian simulations on circuit-based machines based on the block encoding of pseudorandom Hamiltonians.

\section{Protocol}
\label{sec:protocol}

The general structure of a Hamiltonian simulation benchmark on an analog simulator consists of the following steps:
\begin{enumerate}
	\item
	Initialize the device in a state $\ket{\psi_{\text{ini}}}$.
	
	This state, usually a product state, should be easy to prepare
	on the machine.

	\item
	Evolve the state under a simple Hamiltonian.
	
	It should be possible to implement the Hamiltonian, or a sufficiently accurate approximation thereof, on the machine.
	Moreover, its time evolution must either be analytically solvable or be efficiently simulable on a classical computer, with a computational cost that is polynomial in the number of qubits.

	\item
	Stop the evolution at a time $t_*$, perform measurements on the experimental state $\rho_{\text{exp}}^*$, and compare its properties with those of the target state $\ket{\psi_{\text{target}}}$.
	
	The measurements should provide enough information about the experimental state to perform this comparison.
	The target state must be chosen such that all properties of interest can be computed.

	\item Define a score based on the minimum required agreement between the selected properties of the two states.
\end{enumerate}

We discuss each point in turn, starting with the Hamiltonian.
As we will see, the proposed protocol quantitatively evaluates the accuracy of the results. It tests not only whether the device qualitatively reproduces the spread of quantum correlations generated by interactions between the qubits, but also the propagation speed and magnitude of those correlations. Consequently, the final score is sensitive to errors at every stage: state initialization, control of the Hamiltonian parameters, and measurement of the final entangled state.

We focus on machines that are currently, or will soon be, commercially available; this practical scope imposes more restrictions than are typically considered in the research literature.

\subsection{Description}

The protocol we introduce is summarized in \cref{fig:summary}.

\paragraph{Hamiltonian}

The Ising model in a transverse field is one of the simplest and most studied lattice quantum many-body Hamiltonians. It reads
\begin{equation}
    \label{eq:H-ising}
	\begin{aligned}
		H_{\text{Ising}}
			&
			= J \sum_{\braket{i, j}} \sigma^z_i \sigma^z_j
			+	g J \sum_i \sigma^x_i,
	\end{aligned}
\end{equation}
where the spins live on a $d$-dimensional lattice, 
$J$ is the coupling constant between spins,\footnote{In the Rydberg atom platforms discussed below, a convenient energy unit for this coupling is \unit{rad / \micro s}.} $g$ is the strength of the magnetic field in units of $J$, and $\braket{i, j}$ indicates that the sum is performed on pairs of indices $(i, j)$ corresponding to nearest neighbors.
Quantum Ising models have long played an important role in the development of the theoretical tools for many-body physics. 
They have served as a playground to explore numerous phenomena such as phase transitions, quantum criticality or out-of-equilibrium phenomena. 
These paradigmatic models, in dimensions one and two, have also recently been used to study various many-body  dynamics (Trotter, Floquet or kicked) on several quantum devices~\cite{VOVROSH_ConfinementEntanglementDynamics_2021,VOVROSH_SimpleMitigationGlobal_2021,KIM_EvidenceUtilityQuantum_2023,CHERTKOV_RobustnessNearthermalDynamics_2024,ECKSTEIN_LargescaleSimulationsFloquet_2024,SEKI_SimulatingFloquetScrambling_2025,HAGHSHENAS_DigitalQuantumMagnetism_2025}.

The $1d$ version, the so-called quantum Ising chain, has been studied since the 1960s~\cite{katsuraStatisticalMechanicsAnisotropic1962}.
In the absence of a longitudinal magnetic field,
the Hamiltonian \eqref{eq:H-ising} in $1d$
can be mapped to free fermions through the Jordan--Wigner transformation (\cref{app:analytical_surge_time}), implying that most properties---at equilibrium or after a quench---can be computed in polynomial time on a classical computer~\cite{katsuraStatisticalMechanicsAnisotropic1962,mccoyStatisticalMechanicsXY1971,sachdevQuantumPhaseTransitions2000,Calabrese_2012,wuLongitudinalMagnetizationDynamics2020,Mbeng_2024}.

Using tensor network methods, $2d$ systems are much harder to simulate on a classical machine than $1d$ systems, and for this reason it is in $2d$ that we may expect some quantum advantage to be observed for the first time~\cite{VOVROSH_SimulatingDynamicsTwodimensional_2025}.
Nevertheless, due to the rapid entanglement growth, global quench problems are difficult to simulate with tensor networks even in $1d$ (see \cref{app:MPS}). In addition, for a quantum simulator that is capable of realizing Hamiltonians in $1d$ and $2d$, there is no obvious reason why the results would be of better quality in $1d$ than in higher dimensions. For this reason it is also relevant to perform a benchmark with a $1d$ model.
Finally, choosing a $1d$ model ensures the scalability of the benchmark because this model can be solved efficiently.

\paragraph{Rydberg atom simulators}

In its Ising mode, a QPU based on Rydberg atoms implements the following effective Hamiltonian~\cite{ADAMS_RydbergAtomQuantum_2019,henrietQuantumComputingNeutral2020,Morgado:2020:QuantumSimulationComputing,wurtz_aquila_2023}:\footnote{Some platforms offer more general Hamiltonians, but we will focus on the most common model.}
\begin{equation}
	\label{eq:H-rydberg}
	\begin{aligned}
		H_{\text{Rydberg}}
			&
			= \sum_{i < j} \frac{C_6}{\vec r_{ij}^6} \, n_i n_j
				+ \frac{\hbar \Omega(t)}{2} \, \sum_i \sigma^x_i
				\\ & \qquad
				- \hbar \delta(t) \, \sum_i n_i
	\end{aligned}
\end{equation}
where $n_i = (\sigma^z_i + 1) / 2$ is the number operator at site $i$, and $\vec r_{ij} := \vec r_i - \vec r_j$ is the displacement vector between sites $i$ and $j$.
The first sum runs over all pairs of sites $i<j$ and describes the long-range van der Waals spin--spin interaction,
whereas the two remaining sums run over all sites and act as transverse and longitudinal magnetic fields, respectively.
The user can choose the time-dependent laser amplitude $\Omega(t)$ and detuning $\delta(t)$, subject to hardware constraints.

When only nearest-neighbor interactions are retained in \eqref{eq:H-rydberg}, the Hamiltonian reduces to a short-ranged Ising model with transverse and longitudinal magnetic fields.
For an interatomic distance $a$ between nearest neighbors, it is convenient to define the coupling $J := C_6 / (4 a^6)$ and the form of \eqref{eq:H-rydberg} for a ring geometry is given in \cref{app:nn_2_zz}.
Because the long-range interactions decay rapidly as $1 / r^6$, a Rydberg QPU approximates the Ising model quite accurately. For instance, on a sufficiently large ring, the next-nearest-neighbor interaction has strength only $J/64$, compared with $J$ for nearest neighbors.
Long-range interactions can nevertheless produce nontrivial effects in the dynamics of quantum spin models, and of quantum Ising chains in particular, such as anomalous correlation propagation or superballistic behavior~\cite{haukeSpreadCorrelationsLongRange2013,cevolaniUniversalScalingLaws2018,villaUnravelingExcitationSpectrum2019,schneiderSpreadingCorrelationsEntanglement2021}. In the present case, however, the decay exponent is sufficiently large that the model with interactions decaying as $1 / r^6$ is known to behave essentially as a local model~\cite{haukeSpreadCorrelationsLongRange2013}.

We now discuss how to tune $\delta(t)$ and $\Omega(t)$ to bring the Rydberg Hamiltonian \eqref{eq:H-rydberg} into the form of \eqref{eq:H-ising}.
Rewriting $n_i$ in terms of $\sigma^z_i$ shows that \eqref{eq:H-rydberg} contains site-dependent terms that are linear in $\sigma^z_i$ (see \cref{app:nn_2_zz}). In general, these terms cannot be cancelled using a spatially uniform detuning $\delta(t)$.\footnote{They could be cancelled approximately using local addressing, in which case the detuning takes the form $\delta_i(t) = \delta_i \delta(t)$, with $\delta_i \in [0, 1]$.}
Translation-invariant systems are an exception because the longitudinal magnetic field induced by the terms $n_i n_j$ is then site-independent (see \cref{app:nn_2_zz}).
This leads us to consider a $1d$ chain with periodic boundary conditions, with the atoms arranged on a ring at an interatomic distance $a$. This distance determines the interaction energy $J = C_6/(4 a^6)$. Finally, $\Omega$ is fixed in terms of $g$.
The distance $a$, and thus $J$, can be changed without changing $g$. This global change in the energy scale of $H$ can be absorbed into a rescaling of time, leaving the state at the corresponding rescaled time unchanged.
We consider the Ising model at the critical point $g = 1$, where the two terms in the Ising Hamiltonian have similar magnitudes (see \cref{sec:protocol:discussion} for further discussion).

\paragraph{Initial state}

Natural states to start with include the $\sigma^z_i$ and $\sigma^x_i$ eigenstates, respectively $\ket{\downarrow \cdots \downarrow}$ and $\ket{+ \cdots +}$, and the antiferromagnetic state (AFM) $\ket{\downarrow\uparrow \downarrow\uparrow \cdots}$.
These are product states but correlations will develop under the effect of the interaction part of the Hamiltonian.
As discussed in \cref{app:correlations} and \cref{app:impact_dist_score}, we focus
on the global correlation surge: for each of the three initial states mentioned above, the 2-point connected correlations extend across the whole system at a particular time (the \emph{surge time}).

The state $\ket{\downarrow \cdots \downarrow}$ is the natural initial state on Rydberg QPUs.
In principle, other states could be considered, allowing us to test additional state-preparation capabilities of the device.
In particular, the state $\ket{+ \cdots +}$ is interesting because it is the initial state for many quantum algorithms and, from the point of view of Eq.~\eqref{eq:H-ising}, it has a simple theoretical description using the free fermion formalism.\footnote{It has a well-defined fermion parity and, by symmetry, it leads to $\braket{\sigma^z_i(t)} = 0$ at all times.}
However, given the constraints on current machines, it is not possible to prepare other states than $\ket{\downarrow \cdots \downarrow}$ with good fidelity when $a$ is small.\footnote{\label{ft:state-prep}One may try to rotate the state 
	$\ket{\downarrow \cdots \downarrow}$ into $\ket{+ \cdots +}$ using the fields $\Omega(t)$ and $\delta(t)$.	Due to the hardware constraints on the maximal intensity and on the minimal duration of such pulses,
	the rotation cannot be performed arbitrarily fast. On the other hand, the spin-spin interactions cannot be turned off and these interactions will keep modifying the state in a nontrivial way. When the atoms are sufficiently far apart, this effect might be neglected on a short timescale.  But in the regime of interatomic distance considered here, the effect of interactions cannot be neglected during the time needed for the rotation. This therefore prevents the preparation of $\ket{+ \cdots +}$ with a reasonable fidelity.
}
In what follows we will focus on the protocols with $\ket{\psi_{\text{ini}}} = \ket{\downarrow \cdots \downarrow}$ and $\ket{\psi_{\text{ini}}} = \ket{+ \cdots +}$.

\paragraph{Measurements and observables}

Measurements on Rydberg QPUs are done in the $\sigma^z_i$ basis and for all spins at the same time. To access other components, one could perform a global spin rotation just before the measurement, using the fields $\Omega(t)$ and $\delta(t)$. But, when the atoms are close, this operation would suffer the exact same limitation as the preparation of $\ket{\psi_{\text{ini}}} = \ket{+ \cdots +}$ (see \cref{ft:state-prep}). Here, in practice we have to limit ourselves to measurements in the $\sigma^z_i$ basis.
Moreover, the measurement is destructive, which means that it terminates the quantum evolution.
Taken together, these different facts mean that one cannot perform a complete (or even a shadow~\cite{elbenRandomizedMeasurementToolbox2023,notarnicolaRandomizedMeasurementToolbox2023}) tomography of the state, and we cannot use the fidelity as the target metric.

However, in the framework of solid-state physics or quantum chemistry simulations, the purpose of using an analog QPU is not to obtain the entire many-body state. Instead, it can be used to provide estimates of relevant observables for the problem at hand, such as local quantities (energy, particle density, magnetization, etc.) and long-distance correlation functions.
For this reason, we will be interested in 2-point connected correlation functions of $\sigma^z_i$ operators
\begin{equation}
	\begin{aligned}
	g^{(2)}_\ell(t)
		&
		:= \Braket{\sigma^z_0(t) \sigma^z_{\ell}(t)}_c
		\\ &
		:= \Braket{\sigma^z_0(t) \sigma^z_{\ell}(t)} - \Braket{\sigma^z_0(t)} \Braket{\sigma^z_{\ell}(t)},
		\label{eq:g2}
	\end{aligned}
\end{equation}
where $\ell = 1, \ldots, \lfloor L/2 \rfloor$ is the distance from a reference site to the other sites, running from the nearest neighbor to the antipodal site.
\rev{When discussing the raw magnetization data below, we use the notation $g_i^{(1)}(t) := \Braket{\sigma_i^z(t)}$ for the 1-point function, and, when translation invariance is used, its average over equivalent sites.}
Site $0$ is fixed because of translation invariance, and we need to consider only half of the sites because of spatial parity.
The connected 2-point functions vanish by construction in any product state, and it is natural to focus on such a quantity to measure the precision with which the device can reproduce the spread of correlations.

On the QPU, the computation is repeated $n_{\text{shots}}$ times (number of shots, usually a few hundred), each shot yielding a bitstring corresponding to the eigenvalues of the operator $n_i$.
These bitstrings can be used to compute experimental $n$-point correlation functions. The translation invariance can also be exploited to improve the statistics.

\paragraph{Measurement time $t_*$}

\begin{figure}[htp]
	\begin{subfigure}[c]{\linewidth}
		\centering
		\includegraphics[width=\linewidth]{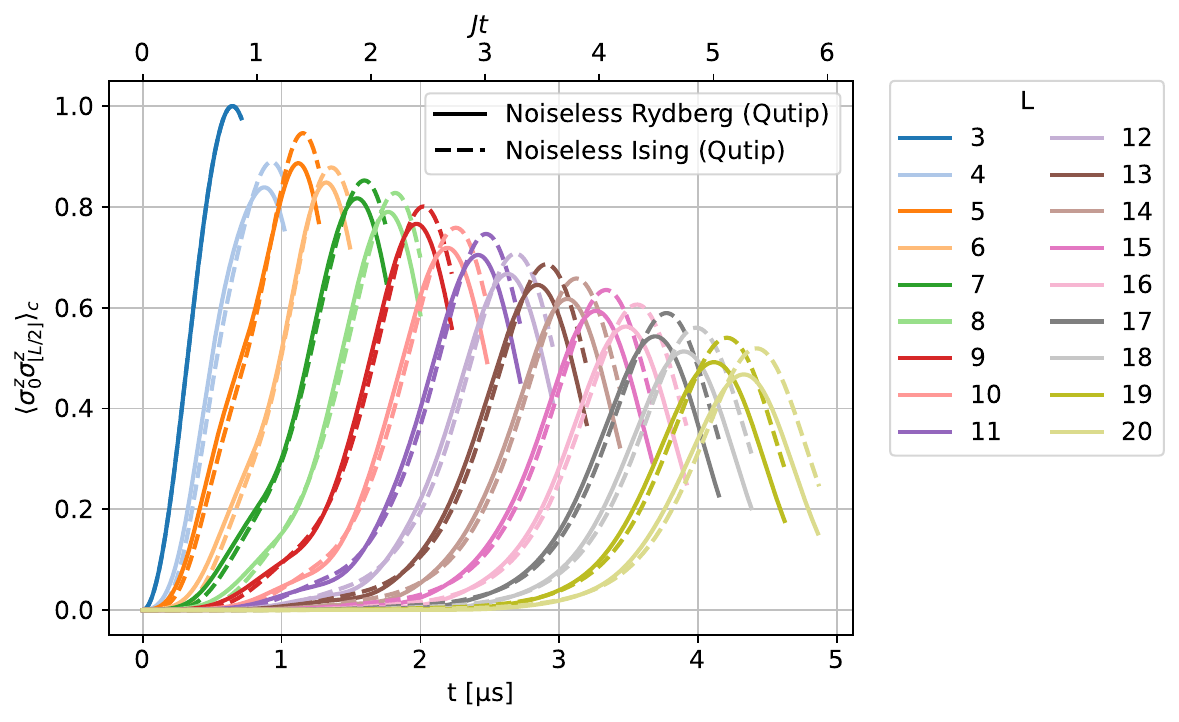}
		\caption{$\ket{\psi_{\text{ini}}} = \ket{+ \cdots +}$}
	\end{subfigure}
	
	\begin{subfigure}[c]{\linewidth}
		\centering
		\includegraphics[width=\linewidth]{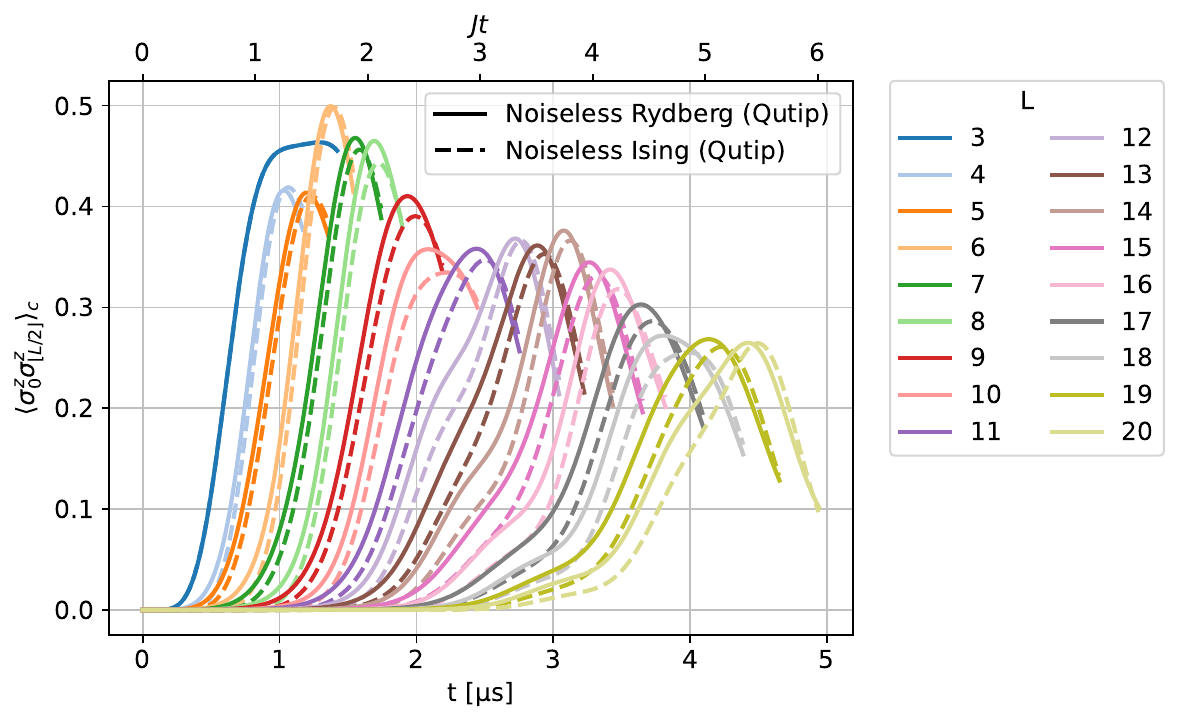}
		\caption{$\ket{\psi_{\text{ini}}} = \ket{\downarrow \cdots \downarrow}$}
	\end{subfigure}
	
	\caption{
		Antipodal 2-point connected correlation functions $g^{(2)}_{\lfloor L/2 \rfloor}(t)$ for the Ising and Rydberg models with $a = \qty{7.5}{\micro m}$ ($J \approx \qty{1.22}{rad \cdot \micro s^{-1}}$), $g = 1$, $L \in [3, 20]$.
		The time evolution is displayed up to $1.1 \times t_*(L)$.
	}
	\label{fig:ising:2pt}
\end{figure}

\begin{figure}[htp]
	\centering
	\begin{subfigure}[c]{\linewidth}
		\centering
	    \includegraphics[width=\linewidth]{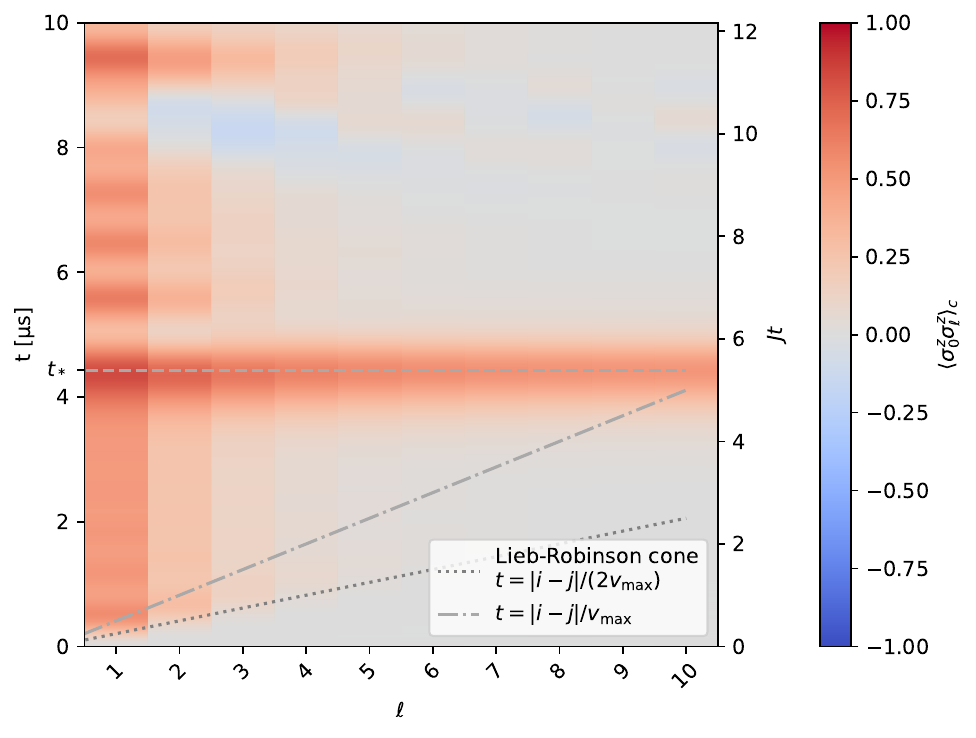}
		\caption{$\ket{\psi_{\text{ini}}} = \ket{+ \cdots +}$}
	\end{subfigure}
	
	\begin{subfigure}[c]{\linewidth}
		\centering
	    \includegraphics[width=\linewidth]{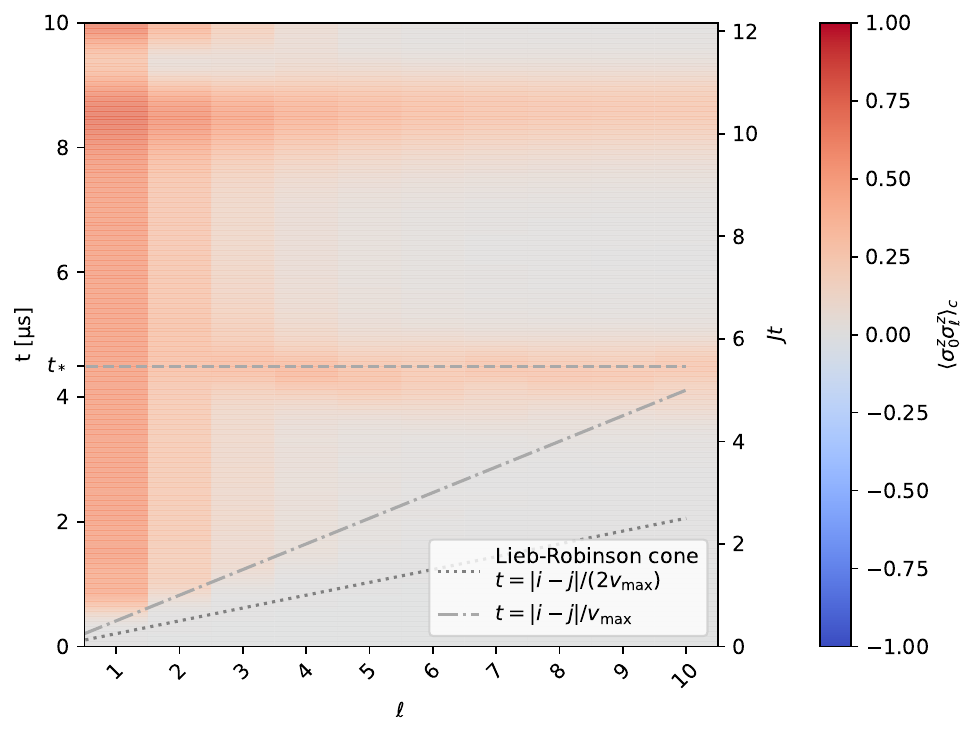}
		\caption{$\ket{\psi_{\text{ini}}} = \ket{\downarrow \cdots \downarrow}$}
	\end{subfigure}
	
	\caption{Spacetime plot of 2-point connected correlation functions $g^{(2)}_\ell(t)$ for the Ising model with $a = \qty{7.5}{\micro m}$ ($J \approx \qty{1.22}{rad \cdot \micro s^{-1}}$), $g = 1$, $L = 20$.
	   (a) initial state $\ket{+\cdots+}$
	   (b) initial state $\ket{\downarrow\cdots\downarrow}$.
	    See \cref{fig:2pt-corr-all-plus} for another representation of this information.
	    \rev{The dashed lines indicate propagation at the maximal quasiparticle group velocity defined in \eqref{eq:max-velocity}; they are not fitted to the plotted correlation data.}
    }
	\label{fig:ising:propagation}
\end{figure}

The final question is to determine the time~$t_*$.
As mentioned in the introduction, by studying the time evolution of the 2-point correlation functions $g^{(2)}_\ell(t)$, we find that there are times when the functions peak for most $\ell$ at the same time (\cref{fig:ising:2pt,fig:ising:propagation})---a phenomenon that we call global correlation surge.
We thus define $t_*$ to be the first of those peaks, and call it the \emph{surge time}.
The peaks in fact come back periodically: $t_*$ essentially corresponds to the time it takes for the information to propagate along half of the ring.
The properties of the system at the surge time are studied analytically in \cref{app:analytical_surge_time}.
\rev{Although the standardized protocol below uses the critical point $g=1$, the existence of a well-defined surge time is not specific to this point. The analysis in \cref{app:analytical_surge_time}, in particular \cref{fig:t_star_estimate}, shows that the rescaled time of the first maximum depends only weakly on $g$ over the range studied, and the additional Ruby data in \cref{fig:2pt-corr-ruby-full-g07,fig:2pt-corr-ruby-full-g13} illustrate that the same analysis can be performed away from criticality. We nevertheless keep $g=1$ as the reference value of MBQS because it gives a simple and demanding standardized instance.}

\paragraph{Score}

We define the score function $P_2(L)$ as the average relative error between the experimental connected 2-point correlation functions $g_\ell^{(2) \exp}(t_*)$ and the corresponding theoretical values $g_\ell^{(2) \text{th}}(t_*)$ for the Ising model, evaluated at the surge time $t_*$:
\begin{equation}
	\begin{aligned}
	P_2(L)
		&
		:= \frac{1}{\lfloor L/2 \rfloor}
		    \\ & \quad
			\times \sum_{\ell = 1}^{\lfloor L/2 \rfloor}
			\left|\frac{g^{(2) \exp}_\ell(t_*) - g^{(2) \text{th}}_\ell(t_*)}{g^{(2) \text{th}}_\ell(t_*)} \right|.
	\end{aligned}
\end{equation}
We expect $P_2(L) = 1$ for classical random values (vanishing connected correlations) and $P_2(L) = 0$ for perfect results.
Note that one can even have $P_2(L) > 1$ if the observed correlations are very far from the theoretical ones.

\rev{In the spirit of the Q-score~\cite{martielBenchmarkingQuantumCoprocessors2021}, the score $S$ of the machine is given by the largest system size $L$ such that the score function remains below some threshold $\epsilon$ for every size from $4$ up to $L$:}
\begingroup
\begin{equation}
	\label{eq:score}
	\begin{aligned}
	S(\epsilon)
		= \max_{\substack{
			L \in \mathbb{N},\ L \ge 4\\
			P_2(L') \le \epsilon\ \forall L'=4,\ldots,L
		}} L.
	\end{aligned}
\end{equation}
\endgroup
\rev{
The lower bound \(L'=4\) is used as a convention for the smallest system size included in the formal score definition. In practice, assigning a score $S=L$ therefore requires benchmark data for all sizes $4,5,\ldots,L$.}
\rev{The threshold $\epsilon$ specifies the maximum allowed average relative error; for example, $\epsilon = 0.01$ corresponds to an average relative error of \qty{1}{\percent}.}

\paragraph{Summary}

The \textbf{Many-Body Quantum Score protocol} with initial state $\ket{\psi_{\text{ini}}}$ and threshold $\epsilon$, $\mathrm{MBQS}_{\psi_{\text{ini}}}(\epsilon)$, consists of the following steps:
\begin{enumerate}
	\item Set up a chain of $L$ equally spaced spin-$\frac{1}{2}$ particles on a $1d$ ring.

	\item Initialize the register with the state $\ket{\psi_{\text{ini}}}$ (canonical choices: $\ket{+ \cdots +}$ or $\ket{\downarrow \cdots \downarrow}$).

	\item Perform a quench with the Ising Hamiltonian at the critical point $H_{\text{Ising}}(J, g = 1)$ for a duration $t_*(L)$ (surge time).

	\item Perform measurements $\{ \sigma^z_i \}$ and compute the experimental connected 2-point functions $g^{(2) \exp}_\ell(t_*)$ (\cref{eq:g2}).

	\item Compute the score function $P_2(L)$ and compare with the threshold $\epsilon$.
	
	\item Define the score $S$ to be the largest $L$ for which the test passes
	for all sizes $4, 5, \ldots, L$.
\end{enumerate}
The protocol has two free parameters: the initial state and the error threshold $\epsilon$. Allowing these parameters to vary makes it possible to define less demanding benchmarks for near-term devices, although the community may eventually converge on a canonical choice, such as $\mathrm{MBQS}_{+}(0.05)$.

The coupling $J$, or equivalently the interatomic distance $a$ in a Rydberg implementation, is not fixed by the protocol. In the ideal Ising model, $J$ sets the overall timescale
but the state attained at fixed $Jt$ is independent of $J$. The optimal value of $J$ on a physical device depends on the hardware. Users should therefore choose $J$ for their platform and report the selected value.

Scores should be compared only when the initial state and threshold are the same. To display performance over a range of thresholds, we encourage reporting a volumetric plot as a function of $L$ and $\epsilon$ (see \cref{sec:test-ruby}).

The protocol allows error mitigation to be used to improve the score. Although some observables are robust against certain forms of noise~\cite{poggiQuantifyingSensitivityErrors2020,DALEY_PracticalQuantumAdvantage_2022,trivediQuantumAdvantageStability2024,caiStochasticErrorCancellation2024}, mitigation may still be necessary to obtain accurate results. Scores and volumetric plots should therefore be reported both with and without error mitigation, and the mitigation procedure should be described in detail.

\subsection{Discussion}
\label{sec:protocol:discussion}

We now examine the extent to which our protocol satisfies the requirements of a well-designed benchmark (this list is adapted from~\cite{Dai:2019:BenchmarkingContemporaryDeep}, see also~\cite{Proctor:2024:BenchmarkingQuantumComputers,Mesman:2024:QuASQuantumApplication}):
\begin{enumerate}
	\setlength{\itemsep}{1ex}

	\item \textbf{Relevance}:
		The goal of a quantum simulation is not to reconstruct the full many-body state---which would require an exponentially large and intractable amount of information---but rather to extract physically meaningful observables. Our protocol targets connected correlation functions at both short and long distances. These quantities are central in condensed matter physics: they distinguish different phases of matter, quantify the system's response to external perturbations, and reveal deviations from mean-field or product-state descriptions.
        In addition, making progress on these many-body problems typically requires high accuracy for the relevant observables. The
        \rev{proposed benchmark addresses this point since the correlations to be measured decrease in amplitude with the system size (see \cref{app:impact_dist_score} and \cref{app:peak_large_systems}). Hence, achieving a larger score not only requires more qubits and a longer evolution time (or a larger circuit depth) but also a larger number of measurement shots to attain the necessary statistical precision.}

	\item \textbf{Representativeness}:
	    Many-body simulation is a central target application for QPUs, with an extensive literature supporting its relevance~\cite{GEORGESCU_QuantumSimulation_2014,BAUER_HybridQuantumClassicalApproach_2016,browaeys_many-body_2020,schollQuantumSimulation2D2021,altmanQuantumSimulatorsArchitectures2021,DALEY_PracticalQuantumAdvantage_2022,andersenThermalizationCriticalityAnalogue2025,SCHLEICH_CrackingChemistryQuantum_2025,JULIA-FARRE_HybridQuantumclassicalAnalog_2025}.
	    Our metric is based on a well-established model in many-body physics, featuring one of the simplest Hamiltonians that nonetheless exhibits rich, nontrivial behavior.

	\item \textbf{Equity}:
		Since the target (Ising) Hamiltonian is particularly simple, the protocol can in principle be implemented on a wide range of platforms and one should be able to fairly compare different devices.

	\item \textbf{Repeatability}:
		The score is completely determined by two parameters only (the choice of the initial state and the threshold $\epsilon$) and the results can be verified by comparing experimental data with analytic computations.

	\item \textbf{Cost-effectiveness}:
			The tests are resource-efficient, requiring measurements only at a single time $t_*$. They do not rely on full-state tomography, which would be prohibitively expensive. However, since the global correlation surge to be detected decreases in amplitude with system size (\cref{app:peak_large_systems}), the number of shots needed to resolve it will increase with $L$.
		
	\item \textbf{Scalability}:
		The reference results can be computed for practically any system size  because correlation functions in the $1d$ Ising model can be computed in polynomial time
		($L$ of the order of a few thousand can be treated on a laptop).

	\item \textbf{Transparency}:
		The metrics can be easily compared, and they are also informative for non-experts since they describe the amounts of quantum correlations which can be correctly reproduced by the QPU.
\end{enumerate}

\begin{figure}
	\centering
	\includegraphics[width=\linewidth]{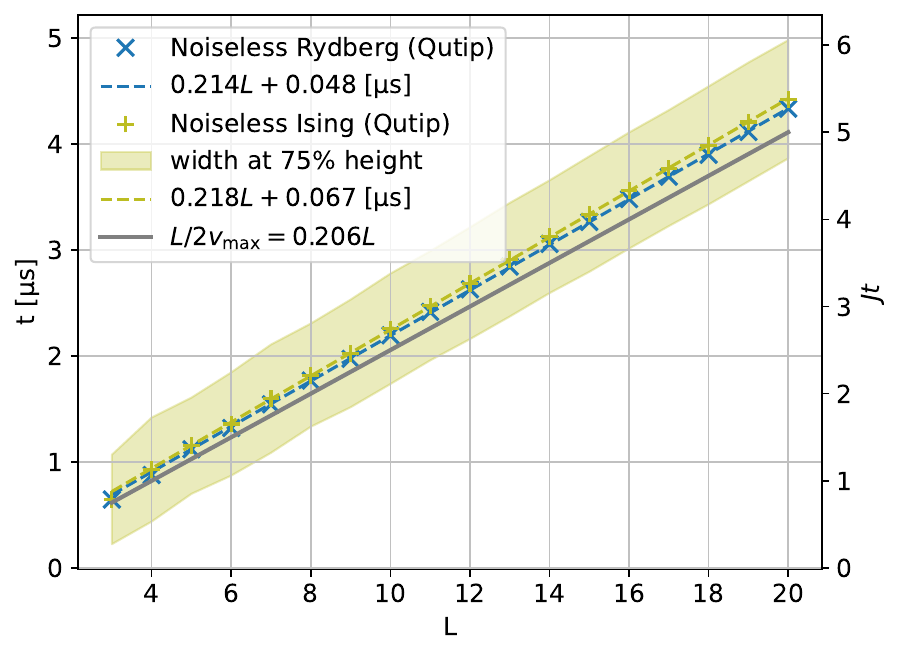}
	
	\caption{Surge times for the Ising and Rydberg models with $g = 1$, $a = \qty{7.5}{\micro m}$ ($J \approx \qty{1.22}{rad \cdot \micro s^{-1}}$), $\ket{\psi_{\text{ini}}} = \ket{+ \cdots +}$.
	    The cross and plus signs mark the values computed by finding the maximum of the antipodal 2-point connected correlation functions.
	    The linear regressions are indicated by lines ($R^2 = \num{0.99996}$ for both), and the yellow shaded area indicates the width of the peak at \qty{75}{\percent} height.
	    Finally, we also show the time computed from the Lieb--Robinson velocity.
	    }
	\label{fig:peak-time-reg}
\end{figure}

\begin{figure}
	\centering
	\includegraphics[width=\linewidth]{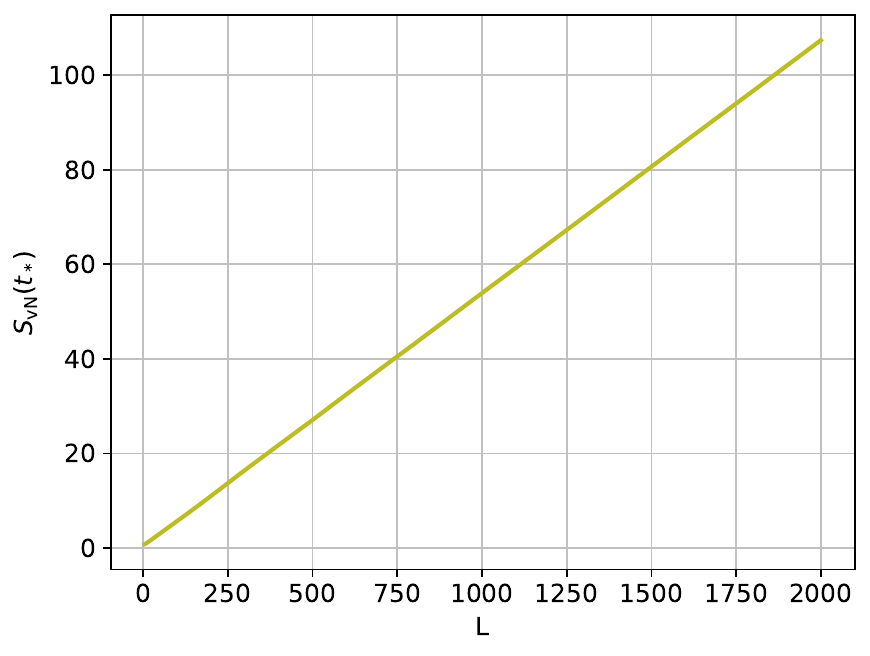}
	
	\caption{Von Neumann entropy for a partition of the chain into two equal parts at the surge time $t_*$, plotted as a function of the system size $L$.
    }
	\label{fig:entropy-L}
\end{figure}

After this general overview of the features of our benchmark, let us examine different aspects of the protocol in more detail and show why it presents multiple challenges for the QPU:
\begin{description}
	\item[Ising model]
			We have chosen the Ising model (with nearest-neighbor interactions) for its simplicity and because it is exactly solvable in $1d$ using free fermions (Jordan--Wigner transformation).
		This means that we can compute its properties in polynomial time ($\mathcal O(L^3)$).
		Although this target Ising model  is well approximated by the native Rydberg Hamiltonian, the two are not identical. This deliberate mismatch demonstrates that analog quantum simulators can be used to emulate Hamiltonians beyond their native model.
	
		\item[Geometry]
		As explained above, we focus on a $1d$ geometry to exploit the integrability of the Ising model. Note however that a $1d$ array typically cannot accommodate the maximum number of atoms supported by the device (current machines constrain the atoms to lie within a disk of approximately $\qty{70}{\micro m}$ in diameter).
		\rev{The choice of a $1d$ geometry should be viewed primarily as a matter of standardization and verifiability, and it is not an attempt to capture the full range of physics accessible on Rydberg platforms.
		$2d$ Rydberg arrays are a natural and important direction for future benchmark extensions, but a $2d$ version would require either a different classically verifiable reference problem or a different acceptance criterion. The present MBQS intentionally keeps the reference problem $1d$ so that the benchmark remains scalable and independently checkable.}
		
		We have also selected periodic boundary conditions because, on a Rydberg machine, this is the only case where the longitudinal ($\sigma^z$) field in the Ising model (which breaks integrability) can be exactly cancelled using a spatially uniform detuning (see \cref{app:nn_2_zz}).
		
		For open boundary conditions, we can still approximately cancel the longitudinal field inside the chain but not on the boundaries: in this case we would have a quantum Ising model with boundary longitudinal field, which is still (free fermion) integrable~\cite{Campostrini:2015:QuantumIsingChains}.
		However, a $1d$ open chain allows fitting only a small number of atoms in the chamber, which makes it less interesting for scalability.
		
		Periodic boundary conditions have two additional advantages.
		First, this case is harder for matrix-product state (MPS) simulations (see \cref{app:MPS}), so it provides an advantage to quantum devices in the competition with classical tensor network methods.
		Indeed, the bond dimension required to describe with an MPS a state on a periodic chain is expected to be the \emph{square} of the bond dimension required to describe a state with the same amount of local entanglement on an open chain.
		Second, taking advantage of the translation invariance one can improve the statistics on the correlation functions and therefore obtain a better precision with the same number of measurements.
	
	\item[Critical point]
		We have chosen the critical point $g_c = 1$ of the Ising model\footnote{In the presence of long-range $1/r^6$ interactions, the critical point is shifted to $g_c \gtrsim 1$.} because the dynamics is non-trivial
		and this is the point where there is the largest competition between the different terms in the Hamiltonian.
		Moreover, it happens that the interatomic distance is approximately equal to the Rydberg blockade radius.
		This also means that the experiments are sensitive to positional noise.

	\item[Surge time]
	    In general, we do not want to perform the benchmark for an arbitrary evolution time, whether constant or $L$-dependent, since it is not clear that such a choice would be suitable for all $L$. The surge time $t_*(L)$, by contrast, has a clear physical meaning and several interesting properties, making it a natural endpoint for the protocol. At this time, correlations have developed across the entire system, providing a clear signal to measure. Since the surge time increases linearly with $L$ (\cref{fig:peak-time-reg}), the benchmark naturally becomes more demanding as $L$ increases. In particular, it probes increasingly long time evolutions on the device, and hence its sensitivity to decoherence.
	    
	    At the surge time, the state is highly entangled:
	    the von Neumann entropy $S_{\text{vN}}(t_*)$ associated with a partition into two equal parts scales with the system size $\mathcal O(L)$ (\cref{fig:entropy-L}). While it corresponds to a local minimum in time, it still obeys a so-called volume law.
	    Consequently, the target state is a non-trivial quantum state and is expected to be difficult to compute classically using MPS. \Cref{app:MPS} presents the results of such MPS simulations. In particular, it shows that with a maximal bond dimension of the order of $400$ one can resolve the correlation peak up to $L=40$, but it fails to capture the peak for $L=50$.

	\item[Interatomic distance]
	    The distance $a$ between the atoms is not fixed by the protocol and is therefore not a benchmark parameter. This is natural because $a$ determines $J$, which merely sets the overall time scale. Moreover, $J$ depends on the $C_6$ coefficient, which may differ from one machine to another. There is also no universally optimal value of $a$: for a given machine, its choice results from a trade-off between $L$ (to fit the atoms within the available chamber), positional noise (which is reduced for larger $a$), and dephasing noise (since larger $a$ leads to a larger $t_*$, and therefore to stronger decoherence effects). For this reason, we leave this choice to the user, who can tune the atomic positions accordingly.

	\item[Extensions]
	    In the future, as QPUs become more efficient, one could consider extensions of the benchmark to make it more demanding. For example, one could check whether the correct values are reproduced for several subsequent peaks (the second and later peaks are particularly sensitive to noise; see \cref{fig:2pt-corr-ruby}), compare higher $n$-point correlation functions, or combine results obtained from several initial states.
\end{description}

\section{A first implementation}
\label{sec:test-ruby}

We have implemented the $\mathrm{MBQS}_{\downarrow}$ protocol described in \cref{sec:protocol} on Ruby, a neutral atom QPU from the Pasqal company.
The QPU has been recently installed and inaugurated at TGCC (Très Grand Centre de Calcul) in France.
We emphasize that this was the first test of a neutral-atom QPU {\em on-premise}.
We have used  the \texttt{pulser} library~\cite{Silverio:2022:PulserOpensourcePackage} with the \texttt{qutip} backend for the noiseless and noisy exact simulations of the Rydberg model, and the \texttt{qutip} library for the exact simulations of the noiseless Ising model~\cite{Lambert:2024:QuTiP5Quantum}.
To facilitate the implementation and reproducibility of the benchmark, we provide the open-source Python package \texttt{mbqs}~\cite{MBQS_package}, which generates the protocol parameters (including their mapping to a neutral-atom QPU), provides the theoretical reference correlations at the surge time, and computes the MBQS metric and score from experimental samples or correlation functions.
Finally, we have used the \texttt{pulser\_myqlm} binding for Eviden's \texttt{Qaptiva} software~\cite{Eviden:MyQLM} to communicate with the Ruby machine and the classical emulator (QLM40).

\begin{figure}
	\centering
	\includegraphics[width=\linewidth]{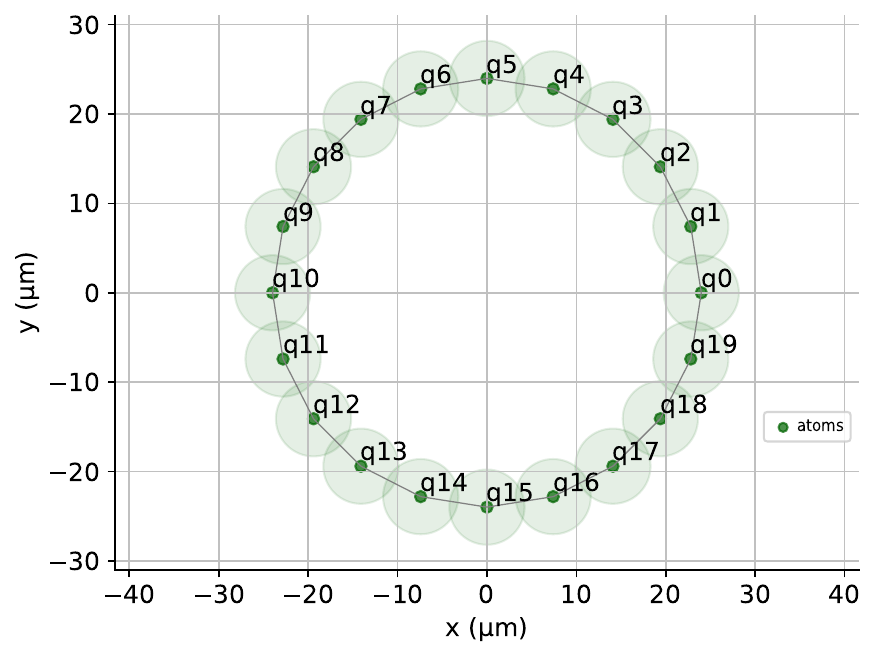}
	\caption{Atom positions, where the circle indicates the Rydberg blockade radius $R_b$.
	}
	\label{fig:atoms}
\end{figure}

The Ruby QPU works with Rubidium atoms with Rydberg level $n = 60$, which gives $C_6 = \qty{865723.02}{\text{rad}\,{\micro m}^6/{\micro s}}$.\footnotemark{}
\footnotetext{For $n = 70$, we have $C_6 = \qty{5420158.53}{\text{rad}\,{\micro m}^6/{\micro s}}$.
}The atoms were placed on a ring with an interatomic distance of $a = \qty{7.5}{\micro m}$
(\cref{fig:atoms}), corresponding to $J \approx \qty{1.22}{rad / \micro s}$.
The laser sequence is built from a single pulse with constant amplitude $\Omega(t) = \qty{2.43}{rad / \micro s}$ (corresponding to $g = 1$) and detuning $\delta(t) = \qty{4.97}{rad / \micro s}$ (corresponding to $m(t) = 0$ in Eq.~\eqref{eq:H-rydberg-ring}).\footnote{Physical devices have a finite modulation bandwidth, which means that the actual sequence realized on the machine is different from the one provided by the user.
    This modulation between the input and output signals can have a strong effect.
    We use the EOM mode (on/off mode) which reduces the effects of modulation for constant pulses~\cite{Pasqal:OutputModulationEOM}.
} With these parameters the blockade radius is $R_b = \qty{8.41}{\micro m}$.

Although the MBQS protocol in principle requires measurements at only a single time, we performed the experiments at several times in order to compare the observed time evolution with the theoretical predictions. 

In practice, the surge times $t_*(L)$ were determined as follows.
We computed the time evolution of the given initial state for the Ising model (without noise).
Next, we determined the locations of the peaks (with nanosecond precision) using \texttt{scipy.signal.find\_peaks()} and additional heuristics to filter out early local maxima from the plateaux. This procedure identifies the first peak displayed in \cref{fig:ising:2pt}.
We then performed a linear regression (see \cref{fig:peak-time-reg} for the plus state) to extrapolate the surge time to higher $L$:
\begin{equation}
    \label{eq:t_star_num}
    J t_*(L)
        \approx \num{0.26} \, L + \num{0.078},
\end{equation}
with $R^2 \approx \num{0.9958}$.
We stored both the exact values determined from the simulations and the linear regression.
Later, when requesting the surge time to run an experiment, we pick the exact value if available, and fall back to the regression otherwise.
A more quantitative understanding of the surge time is provided in \cref{sec:analysis:peak}: however, for the purpose of the experiments, we stick with the method described previously.

\rev{Noisy simulations have been performed with the model described in \cref{app:noise} using the \texttt{QutipBackendV2} with the \texttt{MasterEquation} solver (quantum trajectories over Lindblad equations).
We have used $n_{\text{traj}} = 40$ trajectories.
We thus obtain the mean value of the correlation functions by averaging over the trajectories.
Error bars show the standard error of the mean, computed as the standard deviation of the observable over the trajectories divided by $\sqrt{n_{\text{traj}}}$.}

We have also performed computations with a QLM40 emulator from Eviden, which can run both noiseless and noisy simulations with a high number of qubits.
This allowed us to test the full pipeline, from the job submission to quantum devices within an HPC environment to the data analysis of the bitstrings.
We have gathered $n = \num{2000}$ shots.

\rev{On Ruby, for each system size $L$, we performed $n = \numrange{3000}{9000}$ shots at the surge time $t_*$.
For some system sizes, we have also performed measurements at various fractions of the surge time (from $0.1 t_*$ to $2.1 t_*$), using $n = \numrange{1000}{4000}$ shots.
Next, we have computed the sample mean and standard deviation of the 1-point and 2-point correlation functions.
Error bars again show the standard error of the mean (the sample standard deviation divided by $\sqrt{n}$) and reflect shot noise.}

\begin{figure}
	\begin{subfigure}[c]{\linewidth}
		\centering
		\includegraphics[width=\linewidth]{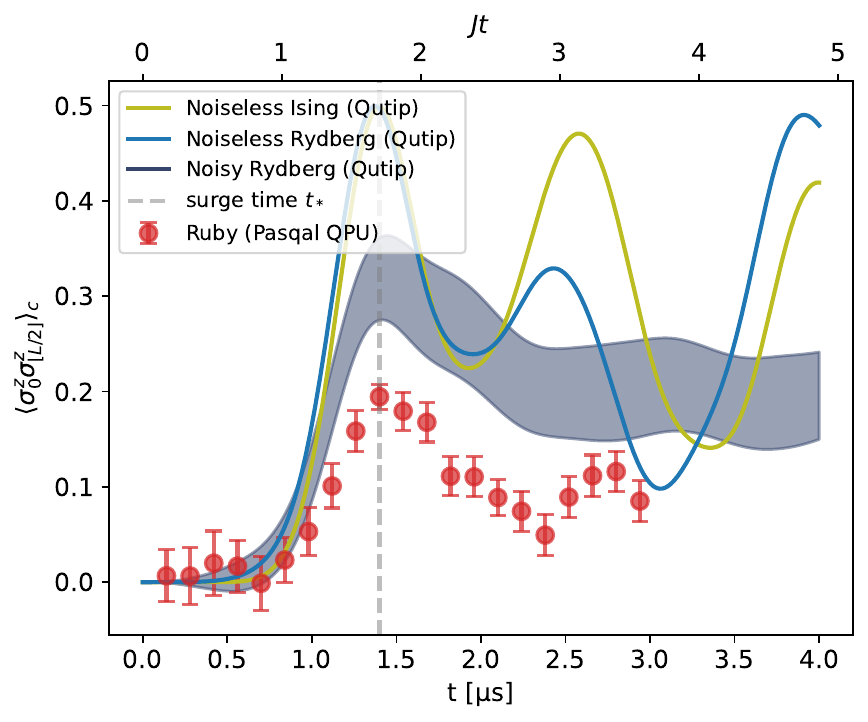}
		\caption{$L = 6$}
	\end{subfigure}
	
	\begin{subfigure}[c]{\linewidth}
		\centering
		\includegraphics[width=\linewidth]{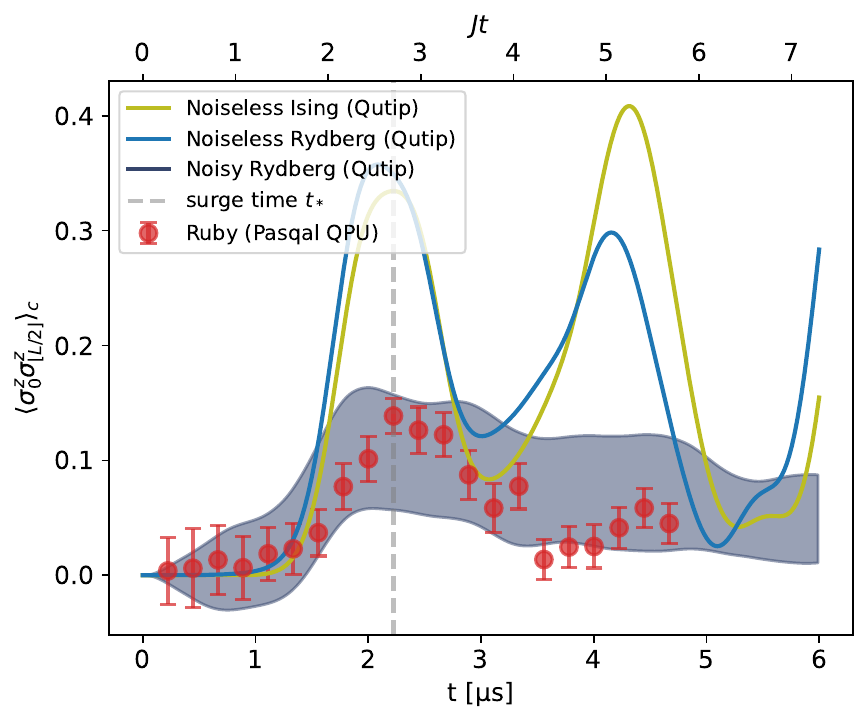}
		\caption{$L = 10$}
	\end{subfigure}
	
	\begin{subfigure}[c]{\linewidth}
		\centering
		\includegraphics[width=\linewidth]{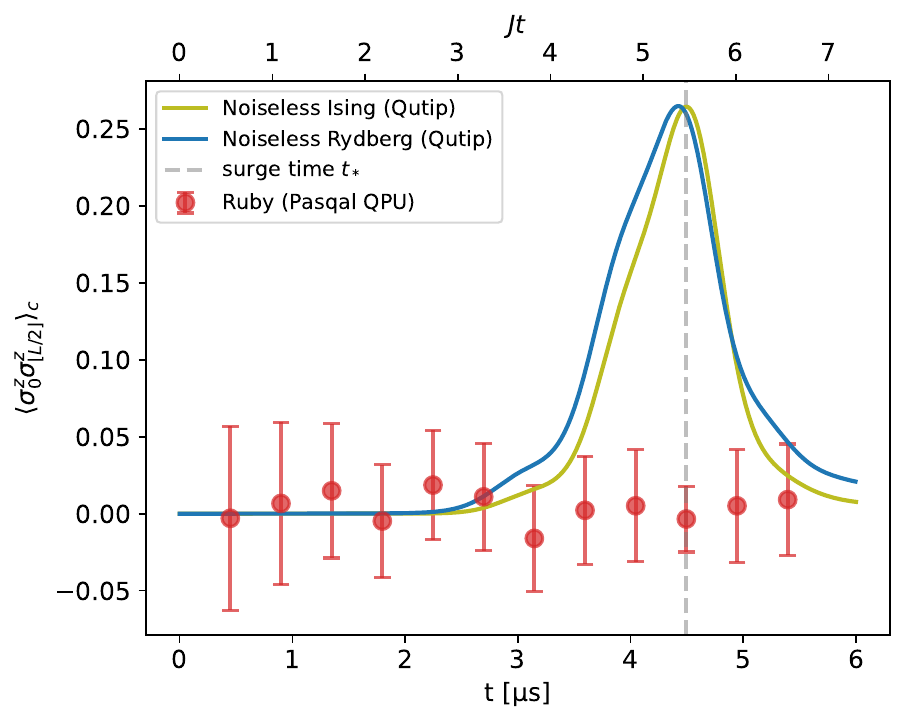}
		\caption{$L = 20$}
	\end{subfigure}
	
	\caption{Antipodal 2-point connected correlation functions for $a = \qty{7.5}{\micro m}$ ($J \approx \qty{1.22}{rad \cdot \micro s^{-1}}$), $g = 1$, $\ket{\psi_{\text{ini}}} = \ket{\downarrow \cdots \downarrow}$.
	    The noise model is described in 
	    \cref{app:noise}.
	}
	\label{fig:2pt-corr-ruby}
\end{figure}

\begin{figure}
	\centering
	
	\begin{subfigure}[c]{\linewidth}
		\centering
		\includegraphics[width=\linewidth]{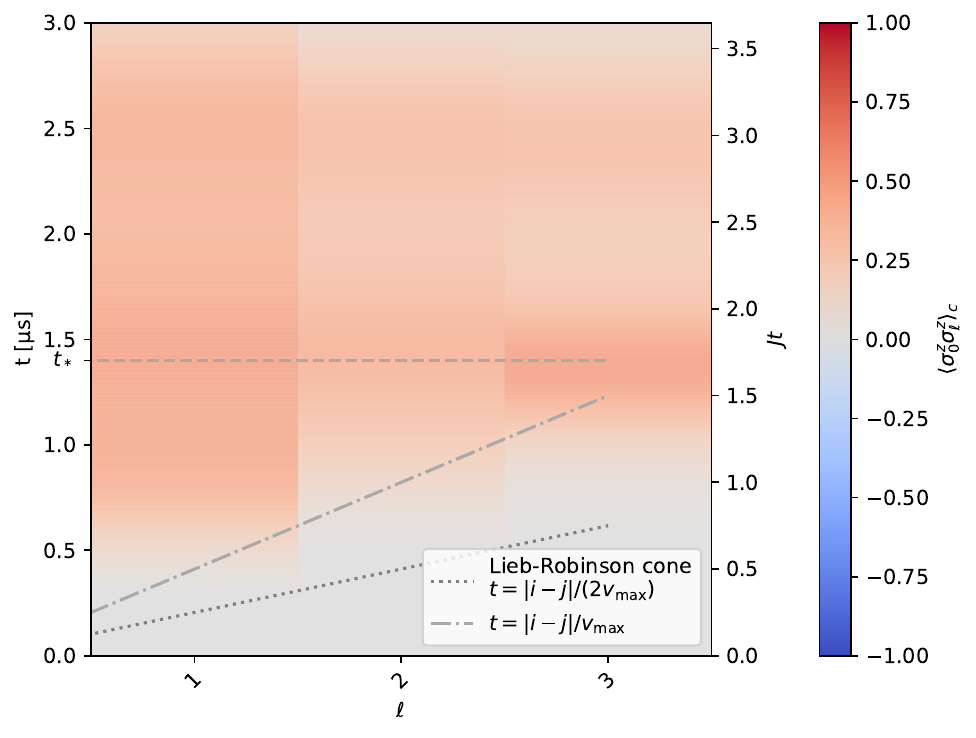}
		\caption{Noiseless Rydberg}
	\end{subfigure}
	
	\begin{subfigure}[c]{\linewidth}
		\centering
		\includegraphics[width=\linewidth]{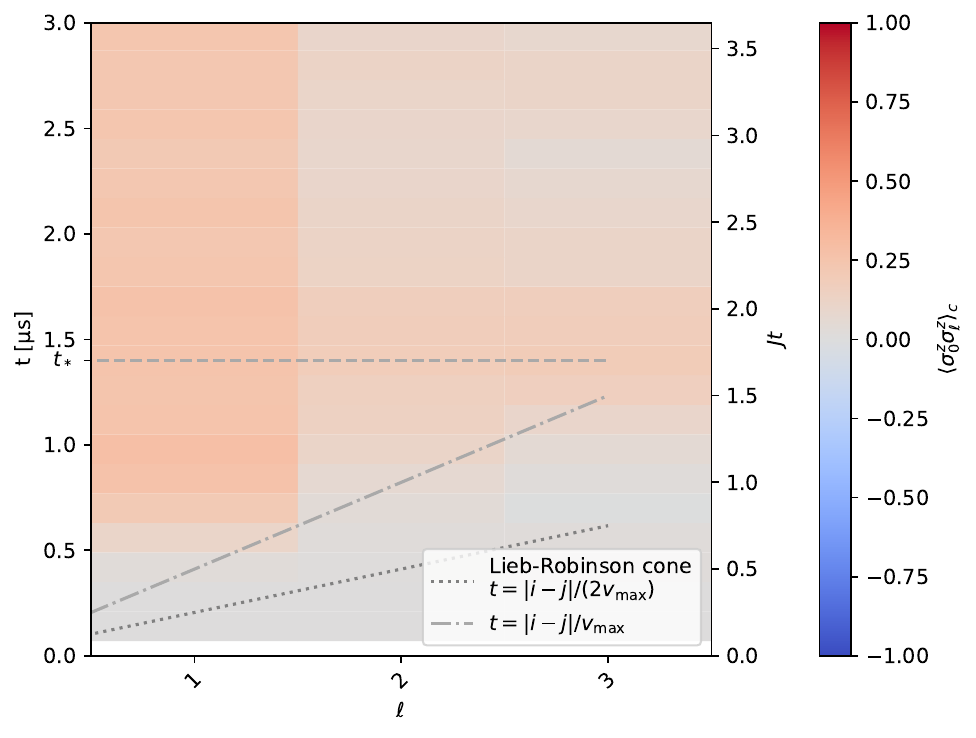}
		\caption{Ruby}
	\end{subfigure}
	
	\caption{Spacetime plots of 2-point connected correlation functions $g^{(2)}_\ell(t)$ for the Rydberg spin chain with $a = \qty{7.5}{\micro m}$ ($J \approx \qty{1.22}{rad \cdot \micro s^{-1}}$), $g = 1$, $\ket{\psi_{\text{ini}}} = \ket{\downarrow \cdots \downarrow}$, $L = 6$.
	    \rev{When shown, dashed propagation lines use the same maximal quasiparticle group velocity as in \cref{fig:ising:propagation} and are not fits to the experimental data.}
    }
	\label{fig:ruby:propagation-L6}
\end{figure}

\begin{figure}
	\centering
	
	\begin{subfigure}[c]{\linewidth}
		\centering
		\includegraphics[width=\linewidth]{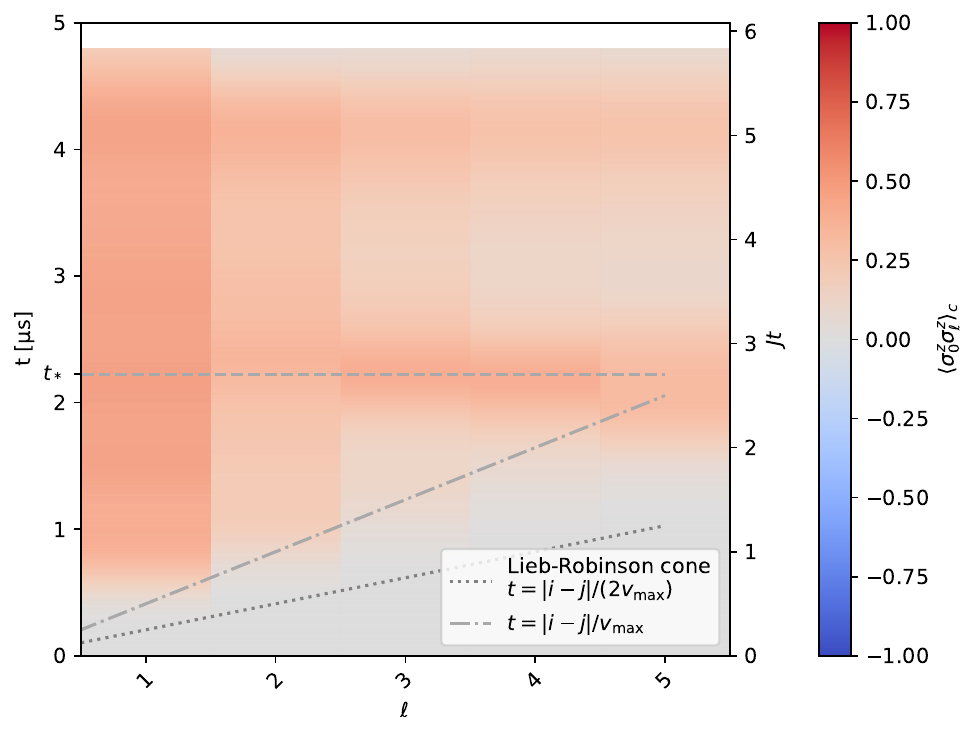}
		\caption{Noiseless Rydberg}
	\end{subfigure}
	
	\begin{subfigure}[c]{\linewidth}
		\centering
		\includegraphics[width=\linewidth]{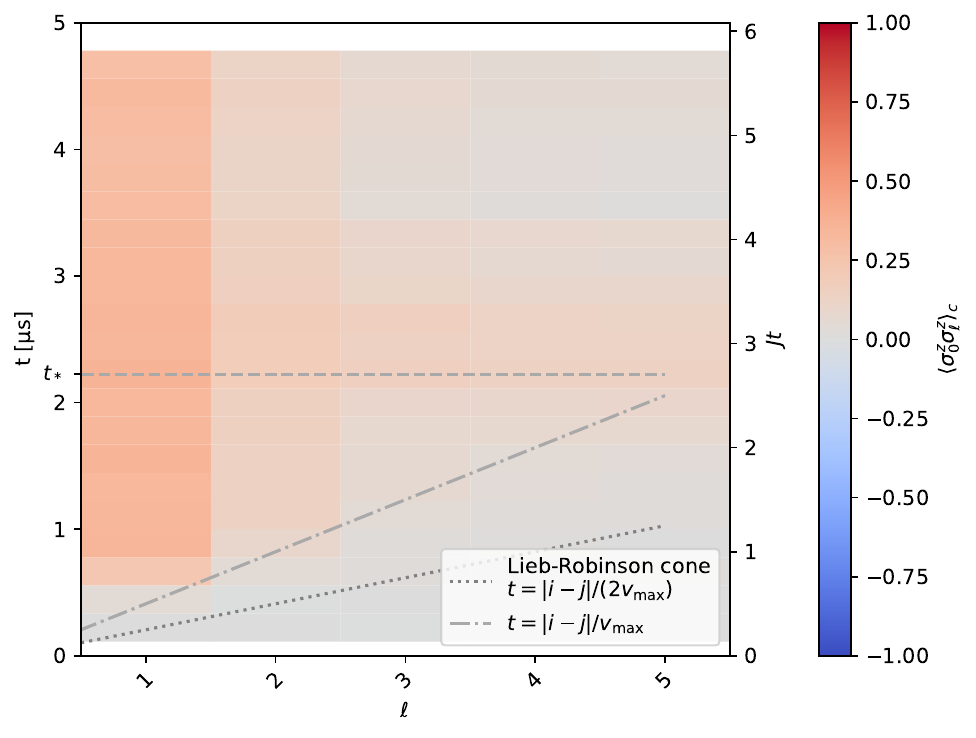}
		\caption{Ruby}
	\end{subfigure}
	
	\caption{Same as \cref{fig:ruby:propagation-L6} for $L=10$.
	    }
	\label{fig:ruby:propagation-L10}
\end{figure}

In \cref{fig:2pt-corr-ruby}, we compare the antipodal 2-point connected correlation functions $g^{(2)}_{\lfloor L/2 \rfloor}(t)$ measured on the machine with the theoretical values for the Ising and Rydberg models.
We also performed noisy simulations for the Rydberg model (see \cref{app:noise} for a detailed discussion) to understand better the observed results.
As we can see, the signal has the overall expected shape but scaled down because of the different noise effects.
We have also performed runs with $L = 20$, but for this size there is essentially no signal left.
The correlation functions at all distances can be compared with the spacetime plots in \cref{fig:ruby:propagation-L6,fig:ruby:propagation-L10}.
While different types of error mitigation can be implemented for analog simulations~\cite{LIU_EfficientlyVerifiableQuantum_2025,Steckmann:2025:ErrorMitigationShot-to-shot}, we only considered readout mitigation.
The formulas are given in \cref{eq:readout-mitigation}.

\begin{figure}
	\centering
	\includegraphics[width=\linewidth]{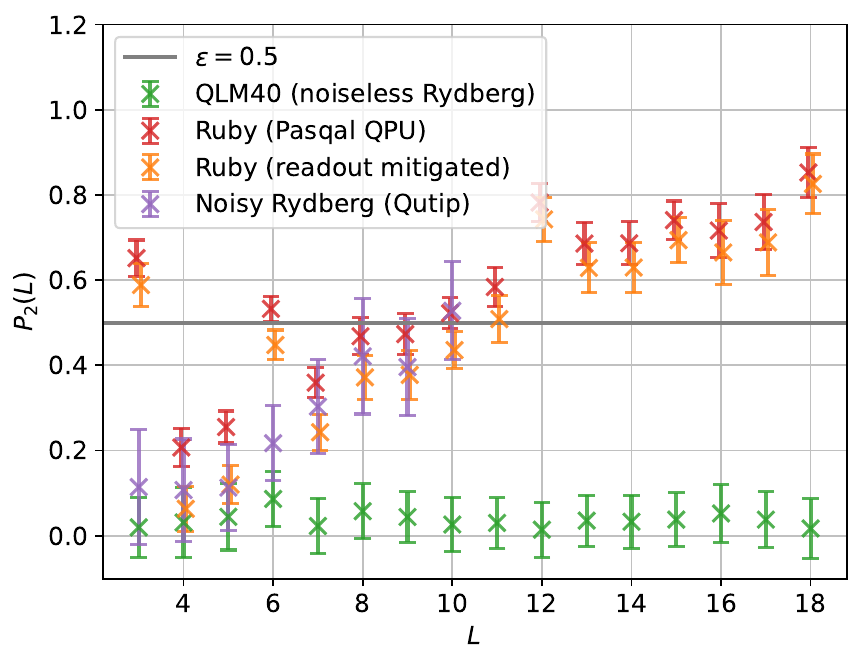}
	\caption{Average relative errors $P_2(L)$ for different devices.
	    QLM40 stands for the noiseless computations performed on Eviden's QLM40 emulator located at TGCC using the software Qaptiva.
	}
	\label{fig:P2}
\end{figure}

\begin{figure}
	
	\begin{subfigure}[c]{\linewidth}
		\centering
		\includegraphics[width=0.65\linewidth]{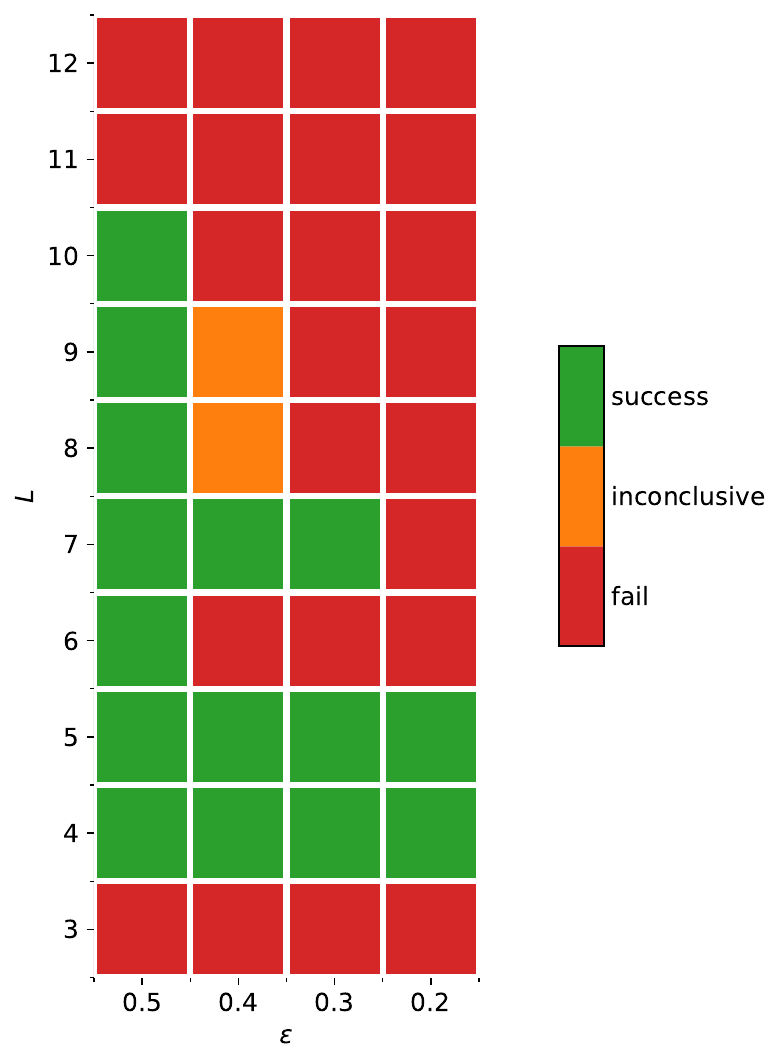}
		\caption{Ruby with readout mitigation.}
	\end{subfigure}
	\vspace*{0.5cm}
	
	\begin{subfigure}[c]{\linewidth}
		\centering
		\includegraphics[width=0.6\linewidth]{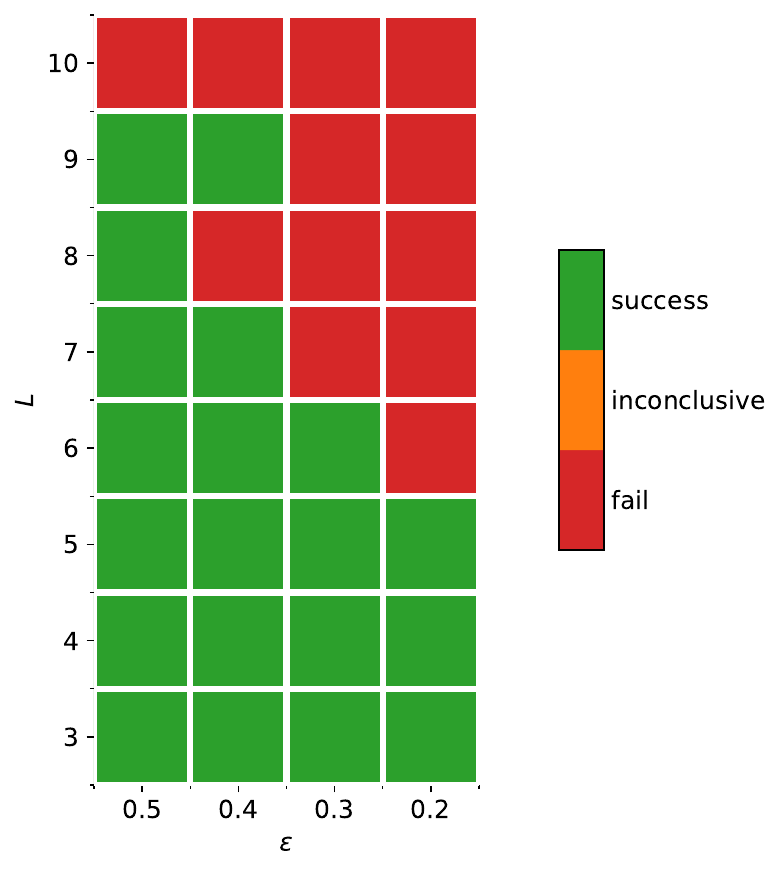}
		\caption{Noisy Rydberg (QuTiP).}
	\end{subfigure}
	\vspace*{0.5cm}
	
	\begin{subfigure}[c]{\linewidth}
		\centering
		\includegraphics[width=0.75\linewidth]{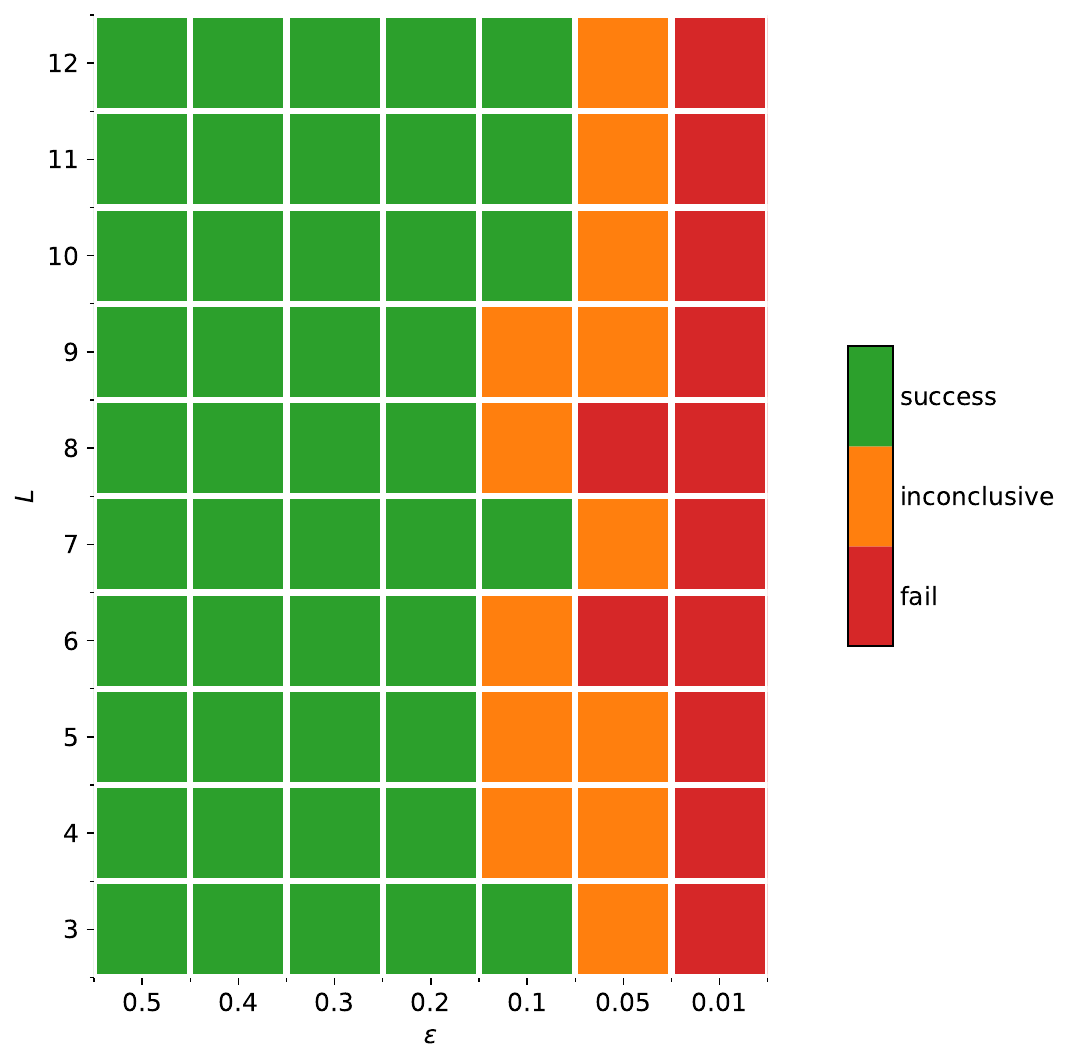}
		\caption{QLM40.}
	\end{subfigure}
	
	\caption{Volumetric plots for different cases.
	    We consider the test unambiguously successful when both the average value and the upper limit of the error bar lie below the threshold, and inconclusive when the average lies below the threshold but the error bar extends above it.
	    The QLM40 plot illustrates what the volumetric plot would look like for a perfect machine with only shot noise.
	}
	\label{fig:volumetric}
\end{figure}

We have gathered the results for different system sizes in \cref{fig:P2},
where the horizontal bar indicates a threshold of $\epsilon = 0.5$ (\qty{50}{\percent} error tolerated).
\rev{For the volumetric plots in \cref{fig:volumetric}, we distinguish three cases: a point is a clear success when the central value and the full error bar lie below the threshold, it is inconclusive when the central value lies below the threshold but the error bar crosses it, and it is a failure otherwise. A score quoted from \cref{fig:P2} should therefore specify whether it is computed from central values only or from this conservative success criterion.}
\rev{Using the central values and the score definition of Eq.~\eqref{eq:score}, which starts at $L=4$, Ruby obtains a score $S = 5$ for the protocol $\mathrm{MBQS}_{\downarrow}(0.5)$ without error mitigation, and the score improves to $S = 10$ with readout mitigation. The unmitigated score is slightly lower than suggested by noisy simulations ($S = 7$), due to the point at $L = 6$, where Ruby performs worse than expected; we do not currently have an explanation for this discrepancy.}
\rev{The point $L = 3$ also performs poorly, which is unexpected because the Hamiltonian for $L = 3$ does not contain long-range interactions and should be among the easiest cases. We therefore regard this point as a useful diagnostic of calibration issues rather than as representative of the scaling trend. Detuning and amplitude calibration errors are likely particularly strong in this case (see \cref{app:noise}).}
Finally, the data for different thresholds can be summarized on volumetric plots (\cref{fig:volumetric}).
We see that we do not expect to be able to use a threshold better than $\epsilon = 0.1$ because of the shot noise.
Comparing the noiseless Rydberg and Ising models shows that they agree within a threshold of $\epsilon = 0.05$, which justifies the use of the Ising model as a reference.

\section{Conclusion}

\rev{In this work, we introduced the Many-Body Quantum Score (MBQS), a scalable, application-level benchmark designed to assess how accurately quantum processors can simulate many-body Hamiltonian dynamics. By focusing on a global quantum quench in the $1d$ transverse-field Ising model and on the accurate reproduction of 2-point connected correlation functions at the surge time, MBQS probes state preparation, Hamiltonian control, coherence, and measurement fidelity within a single protocol. An important property of MBQS is its scalability, which is based on reference results}
\rev{that are classically obtained in polynomial time thanks to the (free fermion) integrability of the $1d$ Ising model.}
\rev{The emergence of a global correlation surge at a time $t_*$ proportional to the system size provides a natural target state and allows one to define a score from the largest size for which correlations are reproduced within a prescribed accuracy threshold.}

\rev{We demonstrated the practical applicability of MBQS through a first test on a commercial neutral atom quantum processor.}
Several studies on Rydberg atom platforms have demonstrated the possibility to tackle many-body problems with tens or even hundreds of atoms~\cite{bernien_probing_2017,schollQuantumSimulation2D2021,ebadiQuantumPhasesMatter2021,bluvsteinControllingQuantumManybody2021,semeghiniProbingTopologicalSpin2021}.
In light of these ongoing efforts to demonstrate quantum advantage, the $1d$  problem considered here may appear relatively simple, and the system sizes modest. However, we emphasize that benchmarking quantum devices on simple and well-controlled models remains essential. In particular, our results show that quantitatively reproducing nontrivial many-body correlations—even between qubits separated by only a few lattice sites—still poses a significant challenge for current quantum technologies.
\rev{Such an analysis, and MBQS in particular, should offer useful tools to monitor, compare, and guide future technological developments in the field of many-body simulations. The present experimental validation concerns one analog neutral atom platform; implementing MBQS on gate-based QPUs and on other analog platforms would provide valuable information concerning the relative strengths and limitations of different approaches to Hamiltonian dynamics.}

\section*{Acknowledgments}

We are grateful to all the people at Pasqal with whom we interacted during the lifetime of the project, in particular Mourad Beji, Antoine Cornillot, Louis-Paul Henry, Henrique Silvério, Vittorio Vitale, and Joseph Vovrosh.
We also acknowledge support from the TGCC, in particular from Marc Joos, Philippe Deniel, and Nicolas Lardjane.
We thank Vincenzo Alba, Edison Carrera, Jacopo de Nardis, Vincent Pasquier, and Kyrylo Snizhko for useful discussions.

This project was provided with HPC resources by GENCI at TGCC thanks to the grant AD010614087 on the supercomputer Joliot Curie's ROME partition.
This work is supported by France 2030 under the French National Research Agency grants No. ANR-22-QMET-0002 (MetriQs-France program and BACQ project~\cite{Barbaresco:2024:BACQApplicationorientedBenchmarks} in particular) and No. ANR-22-EXES-0013, and by the PEPR integrated project EPiQ No. ANR-22-PETQ-0007.

\appendix
\section*{Appendices}
\addcontentsline{toc}{section}{Appendices}

\section{Rydberg model on a ring: from \texorpdfstring{$n_i n_j$ to $\sigma^z_i \sigma^z_j$}{n\_i n\_j to sigmaz\_i sigmaz\_j}}
\label[appsec]{app:nn_2_zz}

The effective Rydberg Hamiltonian for $L$ qubits on a ring is
\begin{equation}
	\label{eq:H-rydberg-ring}
	\begin{aligned}
	H_{\text{Rydberg}}
		&
		=
			J \, \sum_{i=0}^{L-1} \sum_{\ell=1}^{L-1-i} \left(\frac{a}{r_{i,i+\ell}}\right)^6 \, \sigma^z_i \sigma^z_{i+\ell} \\
			& \qquad - J g(t) \sum_i \sigma^x_i
			\\ & \qquad
			+ J m(t) \sum_i \sigma^z_i
			+ C_L(t),
	\end{aligned}
\end{equation}
where the parameters are defined as
\begin{equation}
	\label{eq:rdb:H-ising-param}
	\begin{gathered}
	J
		:= \frac{C_6}{4 a^6},
	\qquad
	g(t)
		:= - \frac{\hbar \Omega(t)}{2 J},
	\\
	m(t)
		:= \hat m
			- \frac{\hbar \delta(t)}{2 J},
	\\
	\hat m
		:= \sum_{\ell=1}^{L-1} \frac{1}{(r_{i,i+\ell} / a)^6}.
	\end{gathered}
\end{equation}
The term $C_L(t)$ is an irrelevant constant.
The distance between atoms $i$ and $i+\ell$ on the ring is
\begin{equation}
	r_{i,i+\ell}
		= a \, \frac{\sin (\pi \ell / L)}{\sin (\pi / L)}.
\end{equation}
Substituting this expression gives the induced longitudinal field
\begin{equation}
	\hat m
		= \frac{L^2 - 1}{945} \, (191 + 23 L^2 + 2 L^4) \sin^6 \frac{\pi}{L}.
\end{equation}
The induced field can be cancelled by choosing a constant detuning $\delta(t) = 2J\hat m/\hbar$, yielding an Ising Hamiltonian without $\sigma^z$ field.

\section{QPU computations and noise model}
\label[appsec]{app:noise}

\subsection{Noise model}

\begin{figure}[htp]
    \centering
    \includegraphics[width=\linewidth]{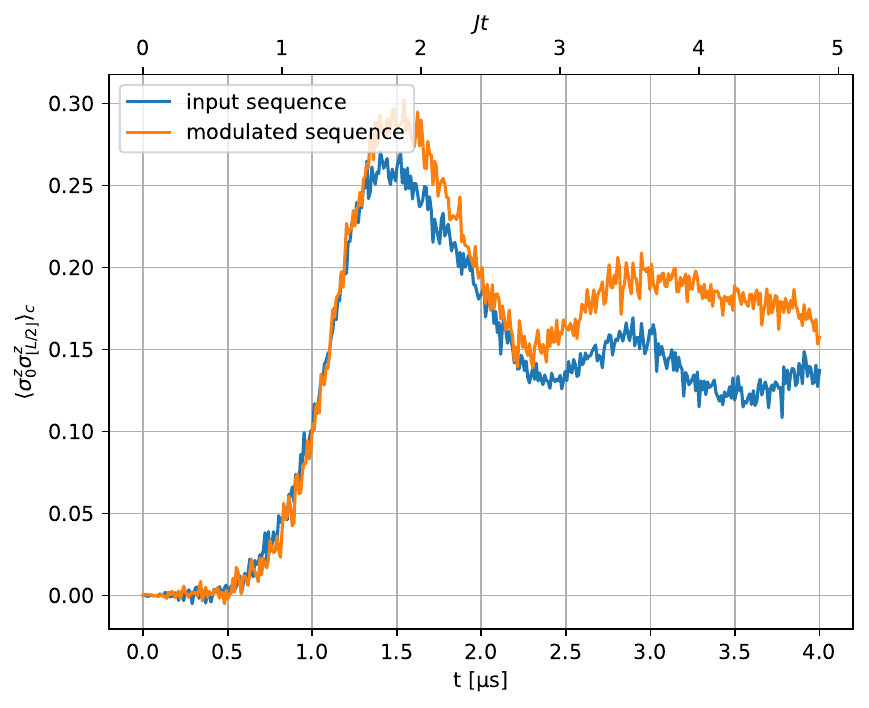}
    \caption{Effect of sequence modulation for the antipodal 2-point connected function on noisy simulations (complete noise model).
        The input sequence corresponds to the sequence requested by the user.
    }
    \label{fig:modulation}
\end{figure}

Neutral atom QPUs are subject to the following sources of noise~\cite{wurtz_aquila_2023,Pasqal:NoiseModel}:
\begin{itemize}
    \item Dissipation due to quantum decoherence:
    \begin{itemize}
        \item relaxation time:
            $T_1 = \qty{100}{\micro s}$
        \item dephasing time:
            $T_2 = \qty{20}{\micro s}$
    \end{itemize}
    
    \item Register noise (denoted by “reg. noise” in the figures) due to uncertainties on the atom positions.
        The atom positions slightly change from shot to shot and this can be modeled by Gaussian noise in the three-dimensional coordinates, with the following standard deviations:
        \begin{equation}
        	\begin{aligned}
        	\sigma^{x,y}_r
        		&
        		= \sqrt{\frac{w_{\text{trap}}^2 T}{4 U_{\text{trap}}}},
        	\\
        	\sigma^z_r
        		&
        		= \frac{\sqrt{2} \pi}{\lambda} \, w_{\text{trap}} \sigma^{x,y}_r,
        	\end{aligned}
        \end{equation}
        where $T$ is the temperature, $U_{\text{trap}}$ the trap depth, $w_{\text{trap}}$ the trap waist, and $\lambda$ the wavelength.
        We have the following values:
        \begin{itemize}
            \item $T = \qty{20}{\micro K}$
            \item $U_{\text{trap}} = \qty{150}{\micro K}$
            \item $w_{\text{trap}} = \qty{1}{\micro m}$
            \item $\lambda = \qty{0.85}{\micro m}$
            \item $\sigma^{x,y}_r = \qty{0.18}{\micro m}$
            \item $\sigma^{z}_r = \qty{0.95}{\micro m}$
        \end{itemize}

    \item State-preparation and measurement (SPAM) errors, which result from missing atoms in the initial state or incorrect measurements at the end of the quantum evolution~\cite{Pasqal:SPAMErrors}.
        Each effect is characterized by a probability:
        \begin{itemize}
            \item preparation error:
                $p_{\text{prep}} = 0.01$
            \item false negative readout:
                $p_{\text{fn}} = 0.07$
            \item false positive readout:
                $p_{\text{fp}} = 0.01$
        \end{itemize}
    
    \item Doppler noise due to thermal motion, which leads to frequency fluctuations.
        This effect is characterized by the temperature~$T$.
    
    \item Laser noise:
    \begin{itemize}
        \item fluctuation of the laser amplitude between shots:
            $\sigma_\Omega = \num{0.02}$
        
        \item finite resolution of the laser beam, whose intensity decreases away from the focus point:
            $w_\Omega = \qty{175}{\micro m}$
        
        \item amplitude and detuning offsets due to the approximate realization of the two-photon transition:
            $\Delta_\Omega$ and $\Delta_\delta$
        
        \item fluctuation of the laser detuning between shots:
            $\sigma_\delta = 2\pi \times \qty{0.05}{rad \cdot \micro s^{-1}}$
        
        \item time-dependent high-frequency detuning noise, defined by its power spectral density (PSD) over the relevant angular-frequency range:
            the values are confidential.
    \end{itemize}
    
    \item Shot noise due to the finite number of measurements that can be performed (a few thousand), which results in a statistical error.
    
    \item Modulation (pulse shaping) of the sequence due to the constraints on how fast the laser values can change~\cite{Pasqal:OutputModulationEOM}.
    This means that there is a difference between the input sequence and the sequence actually run by the machine (\cref{fig:modulation}).
    
\end{itemize}
With the exception of the amplitude offset in \cref{fig:2pt-corr-ruby-full-offset}, we did not adjust any parameter in the noise model.
All parameters were provided by Pasqal and were characterized in separate experiments. The reported values lie within the ranges typical of neutral atom QPUs but may vary over time and between machines.
The characterization was however not performed directly on Ruby, which may explain some deviations; we did not attempt to refine the noise model.
A few noise sources, such as leakage, have been omitted, but they are expected to be subleading compared with the other sources.
The details about the sources of noise and their implementation will be given in an upcoming paper by Pasqal.

We refer to the model used in the simulations, which includes all the effects above except amplitude and detuning offsets, as the “complete noise model”.
\rev{This noise model is used here as a diagnostic tool to interpret the experimental trends; it is not part of the MBQS definition and it is not used to assign the benchmark score. The score is defined directly from the measured correlations and the classically computed reference values.}

\subsection{Discussion of correlation functions}

\begin{figure}
	\begin{subfigure}[c]{\linewidth}
		\centering
    	\includegraphics[width=\linewidth]{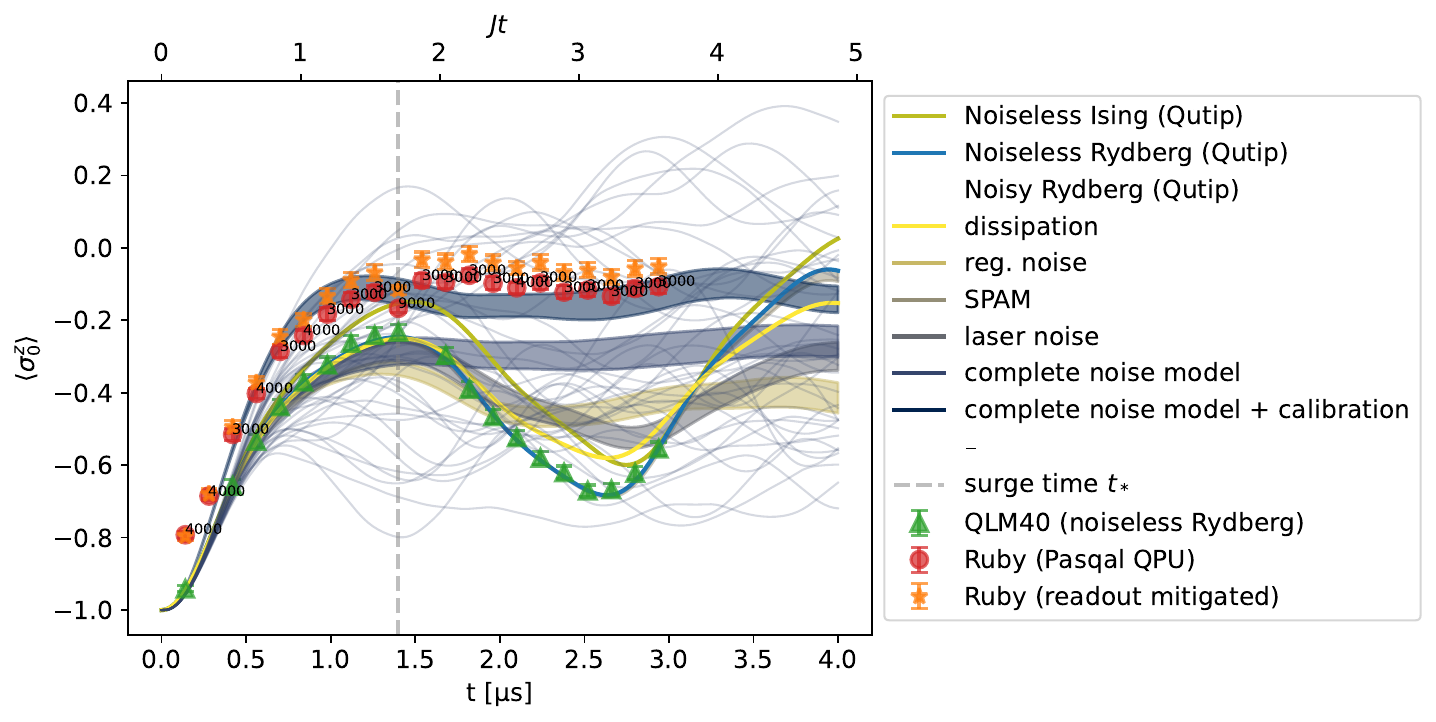}
		\caption{1-point}
	\end{subfigure}
	
	\begin{subfigure}[c]{\linewidth}
		\centering
	    \includegraphics[width=\linewidth]{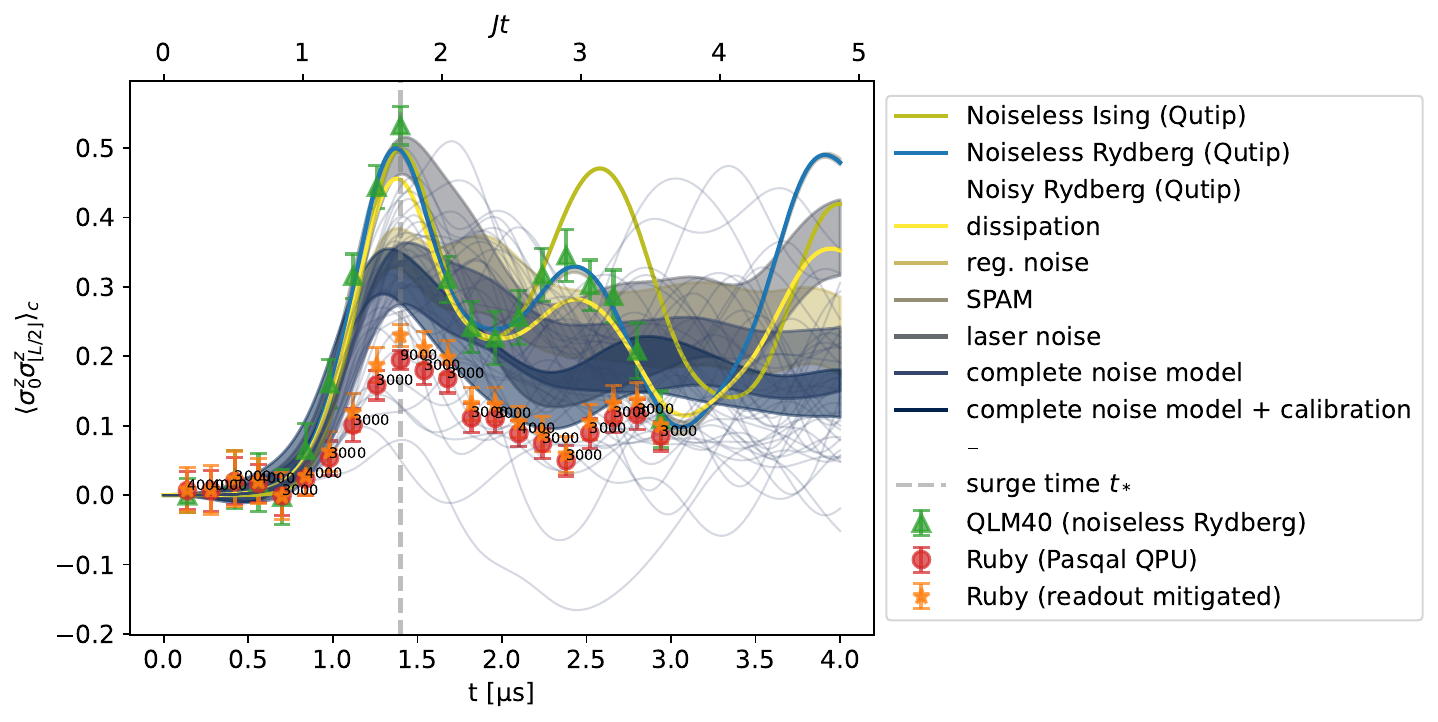}
		\caption{2-point}
	\end{subfigure}
	
	\caption{1- and 2-point correlation functions for $a = \qty{7.5}{\micro m}$ ($J \approx \qty{1.22}{rad \cdot \micro s^{-1}}$), $g = 1$, $\ket{\psi_{\text{ini}}} = \ket{\downarrow \cdots \downarrow}$ and $L = 6$.
	}
	\label{fig:2pt-corr-ruby-full-L6}
\end{figure}

\begin{figure}
	\begin{subfigure}[c]{\linewidth}
		\centering
    	\includegraphics[width=\linewidth]{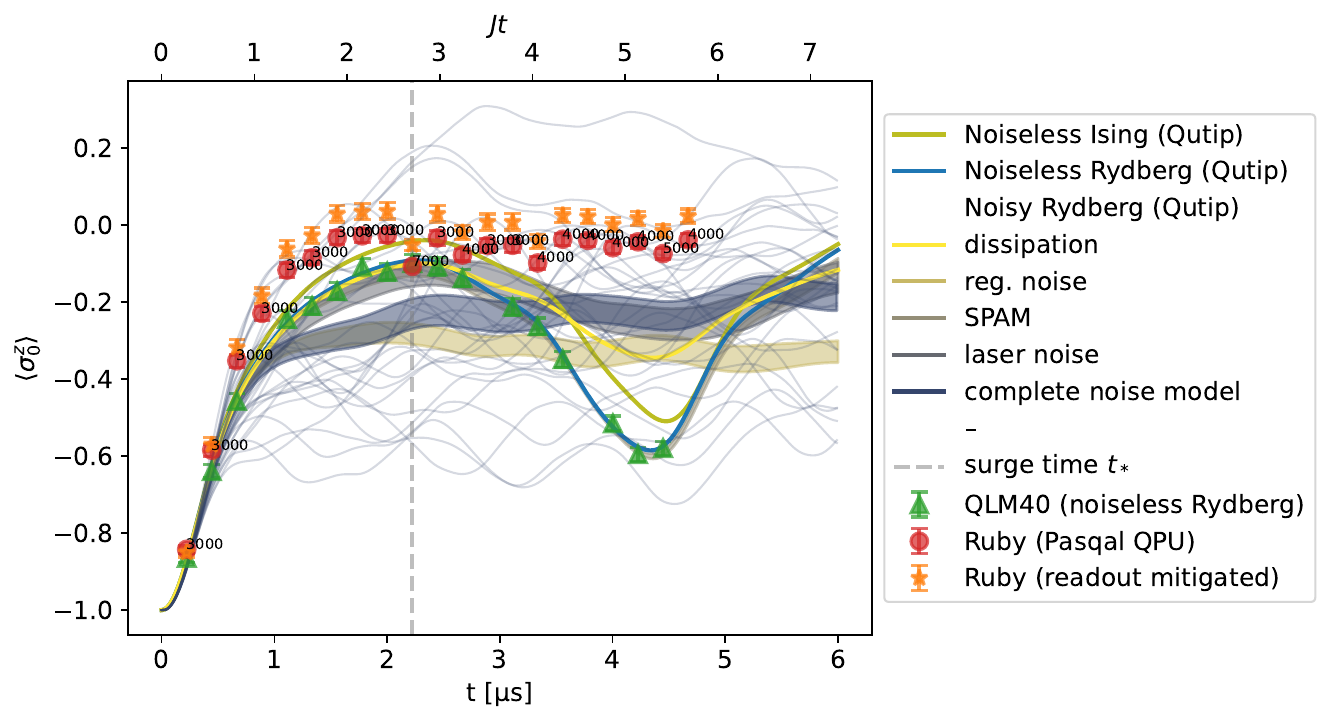}
		\caption{1-point}
	\end{subfigure}
	
	\begin{subfigure}[c]{\linewidth}
		\centering
	    \includegraphics[width=\linewidth]{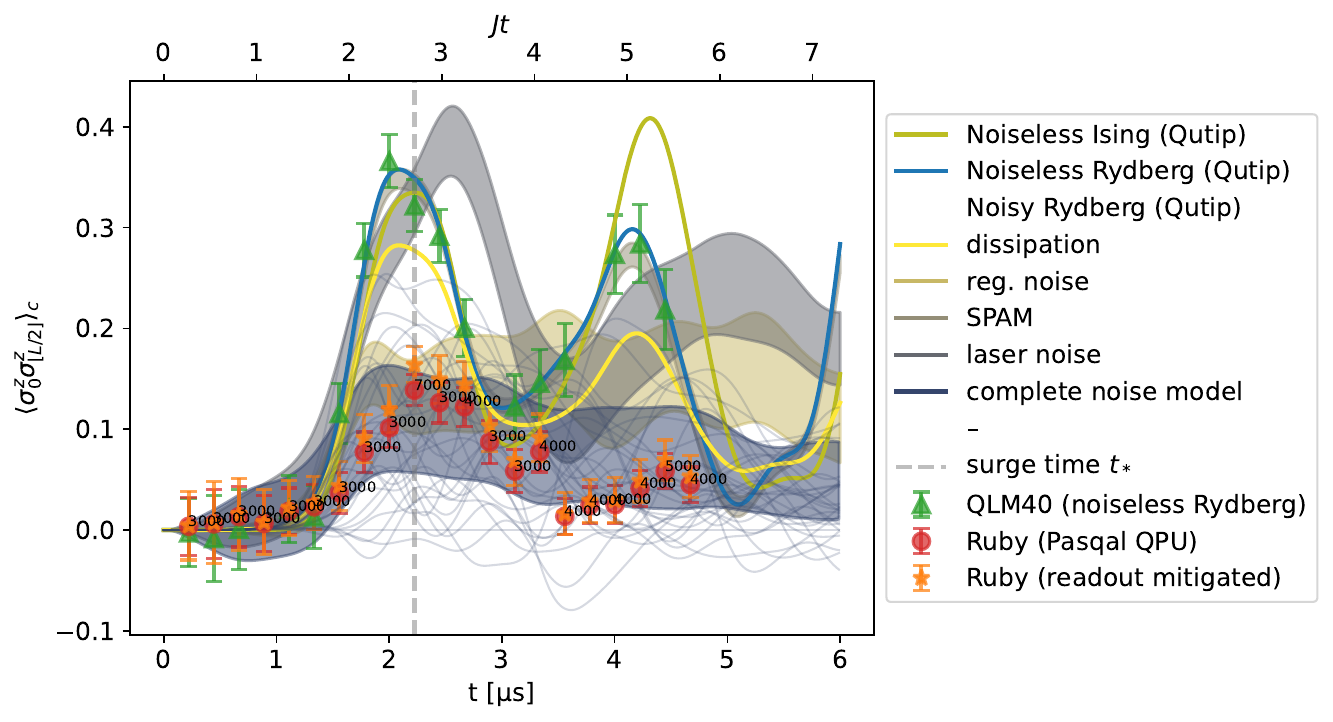}
		\caption{2-point}
	\end{subfigure}
	
	\caption{1- and 2-point correlation functions for $a = \qty{7.5}{\micro m}$ ($J \approx \qty{1.22}{rad \cdot \micro s^{-1}}$), $g = 1$, $\ket{\psi_{\text{ini}}} = \ket{\downarrow \cdots \downarrow}$ and $L = 10$.
	}
	\label{fig:2pt-corr-ruby-full-L10}
\end{figure}

In \cref{fig:2pt-corr-ruby-full-L6,fig:2pt-corr-ruby-full-L10}, we plot the 1-point and antipodal 2-point functions with the different sources of noise, together with the results from Ruby (with and without readout mitigation).
We also display the results for the QLM40 emulation to quantify the effect of shot noise.

A comparison of the noiseless and noisy simulations for the Rydberg model shows that atom disorder is the dominant source of noise, followed by dephasing, SPAM errors (mostly false negatives), and laser errors (mostly high-frequency detuning noise).
Note that dephasing and SPAM errors used to be larger, but recent progress has reduced them (for example, we now have $T_2 = \qty{20}{\micro s}$, compared with $T_2 = \qty{4.5}{\micro s}$ previously).

\begin{figure}
	\begin{subfigure}[c]{\linewidth}
		\centering
    	\includegraphics[width=\linewidth]{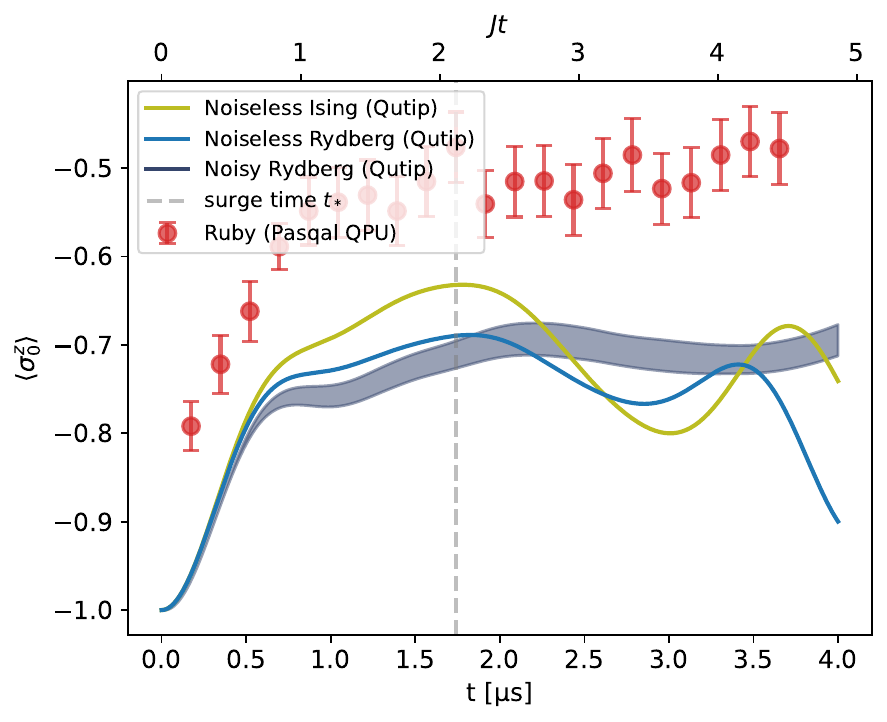}
		\caption{1-point}
	\end{subfigure}
	
	\begin{subfigure}[c]{\linewidth}
		\centering
	    \includegraphics[width=\linewidth]{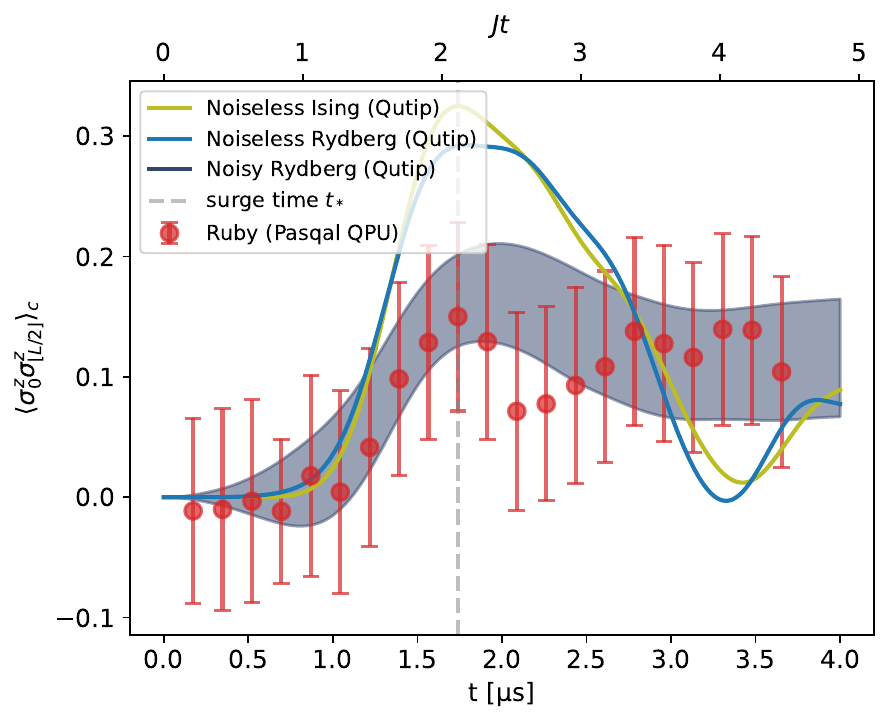}
		\caption{2-point}
	\end{subfigure}
	
	\caption{Correlation functions for $a = \qty{7.5}{\micro m}$ ($J \approx \qty{1.22}{rad \cdot \micro s^{-1}}$), $g = 0.7$, $\ket{\psi_{\text{ini}}} = \ket{\downarrow \cdots \downarrow}$ and $L = 6$ (shots: $n = 500$).
	}
	\label{fig:2pt-corr-ruby-full-g07}
\end{figure}

\begin{figure}
	\begin{subfigure}[c]{\linewidth}
		\centering
	    \includegraphics[width=\linewidth]{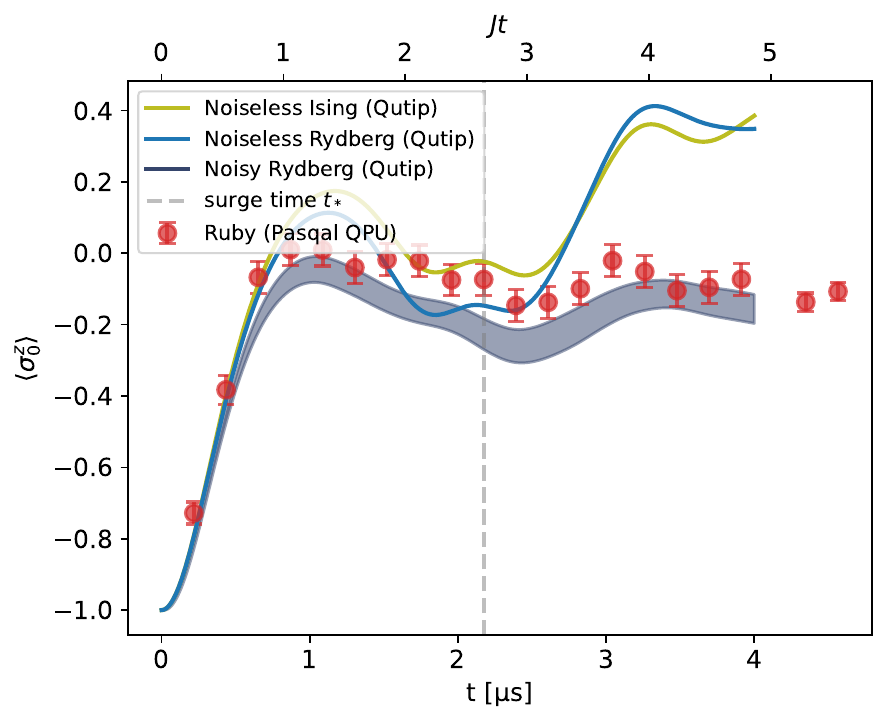}
		\caption{1-point}
	\end{subfigure}
	
	\begin{subfigure}[c]{\linewidth}
		\centering
	    \includegraphics[width=\linewidth]{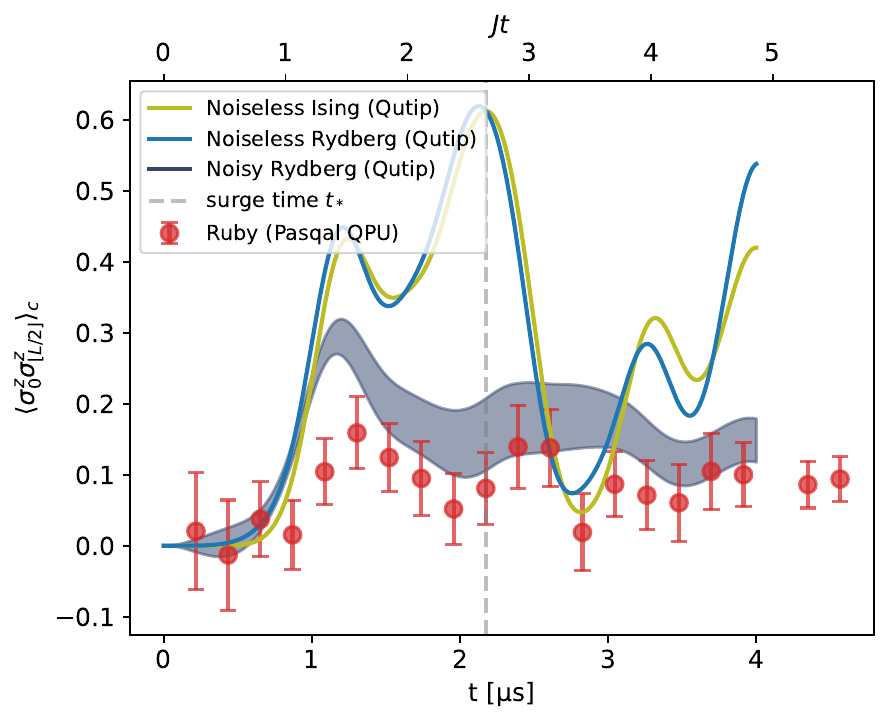}
		\caption{2-point}
	\end{subfigure}
	
	\caption{Correlation functions for $a = \qty{7.5}{\micro m}$ ($J \approx \qty{1.22}{rad \cdot \micro s^{-1}}$), $g = 1.3$, $\ket{\psi_{\text{ini}}} = \ket{\downarrow \cdots \downarrow}$ and $L = 6$ (shots: $n = 500$).
	}
	\label{fig:2pt-corr-ruby-full-g13}
\end{figure}

\begin{figure}
	\centering
	\begin{subfigure}[c]{\linewidth}
		\centering
	    \includegraphics[width=\linewidth]{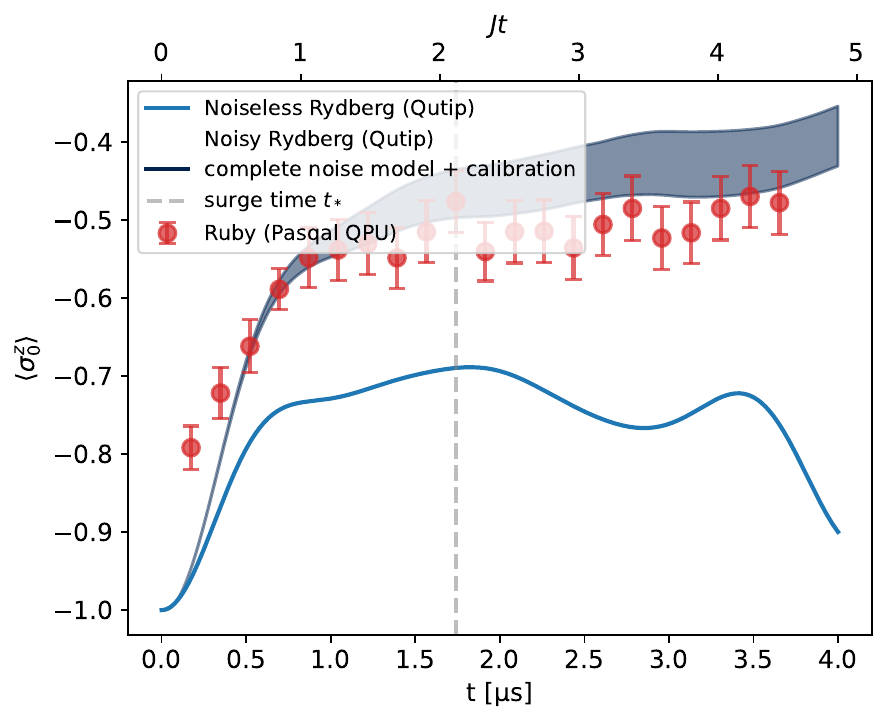}
		\caption{$g = 0.7$}
	\end{subfigure}
	
	\begin{subfigure}[c]{\linewidth}
		\centering
	    \includegraphics[width=\linewidth]{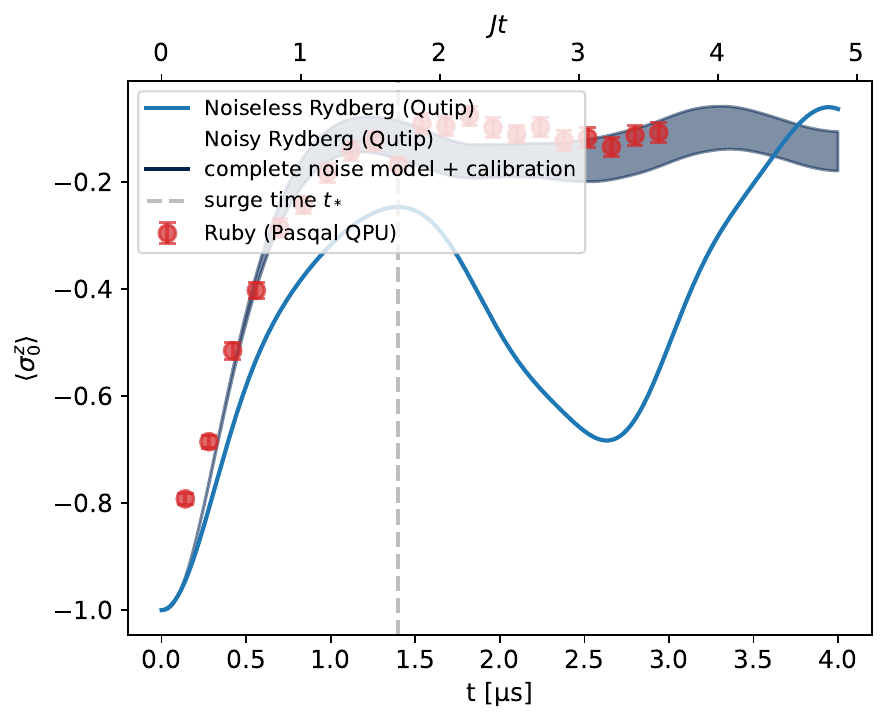}
		\caption{$g = 1.3$}
	\end{subfigure}
	
	\caption{Correlation functions for $a = \qty{7.5}{\micro m}$ ($J \approx \qty{1.22}{rad \cdot \micro s^{-1}}$), $\ket{\psi_{\text{ini}}} = \ket{\downarrow \cdots \downarrow}$ and $L = 6$ with an amplitude offset $\Delta_\Omega = 2\pi \times \qty{0.07}{rad \cdot \micro s^{-1}}$ (shots: $n = 500$).
	}
	\label{fig:2pt-corr-ruby-full-offset}
\end{figure}

We observe that the noise model does not reproduce accurately the experimental 1-point function.
It is not yet completely clear what the origin of the discrepancy is.
\rev{This discrepancy should therefore be interpreted as a limitation of the present device-level noise model, most likely related to calibration or omitted hardware effects, rather than as a limitation of the MBQS protocol itself.}
One could argue that the system close to the critical point is very sensitive to any change of parameters, such that minor calibration errors on the machine would get translated into large deviations in the predictions.
However, we checked that this discrepancy is also present for $g = 0.7$ and $g = 1.3$ (\cref{fig:2pt-corr-ruby-full-g07,fig:2pt-corr-ruby-full-g13}).\footnotemark{}
\footnotetext{Note that this also shows that our protocol makes sense for values $g \neq 1$.
}Moreover, the discrepancy cannot be attributed to an underestimation of the theoretical error bars, since the shift is consistently in the same direction relative to the noisy simulations.
The most likely explanation is an error in the laser calibration, for example due to an imperfect realization of two-photon transitions.
This can be simulated using constant amplitude and detuning offsets; however, different offset values may be needed for different values of the system parameters (such as $g$ and $L$).
We have found that $\Delta_\Omega \approx 2\pi \times \qty{0.07}{rad \cdot \micro s^{-1}}$ brings the simulations closer to the 1-point function, especially at early times (see \cref{fig:2pt-corr-ruby-full-offset}), but it also slightly reduces the agreement for the 2-point function.

\subsection{Readout mitigation}

False positive and negative errors on readout can easily be mitigated by rescaling the correlation functions as follows:
\begin{equation}
    \label{eq:readout-mitigation}
    \begin{aligned}
    \braket{\sigma^z_i}^{\text{mit}}
    	&
    	= \frac{\braket{\sigma^z_i} - d}{1 - s},
    \\
    \braket{\sigma^z_i \sigma^z_j}_c^{\text{mit}}
    	&
    	= \frac{\braket{\sigma^z_i \sigma^z_j}_{c}}{(1 - s)^2},
    \end{aligned}
\end{equation}
where
\begin{equation}
    s = p_{\text{fp}} + p_{\text{fn}},
    \qquad
    d = p_{\text{fp}} - p_{\text{fn}}
\end{equation}
and we assumed $i \ne j$.

\subsection{Data}

Let us comment briefly on the data we collected.
We have repeated the measurements several times for different durations and system sizes (in general, we have $n = 1000$ shots per experiment).
We observed that the results vary slightly from one day to another: for this reason, for a given size, we kept data from the same day.
Moreover, we found that, in some cases, the machine yields vanishing correlation functions; we removed these data because this behavior seems to indicate that the machine was not in a healthy state.

\section{Global correlation surge for different initial states}
\label[appsec]{app:correlations}

\begin{figure}
	\begin{subfigure}[c]{\linewidth}
		\centering
    	\includegraphics[width=\linewidth]{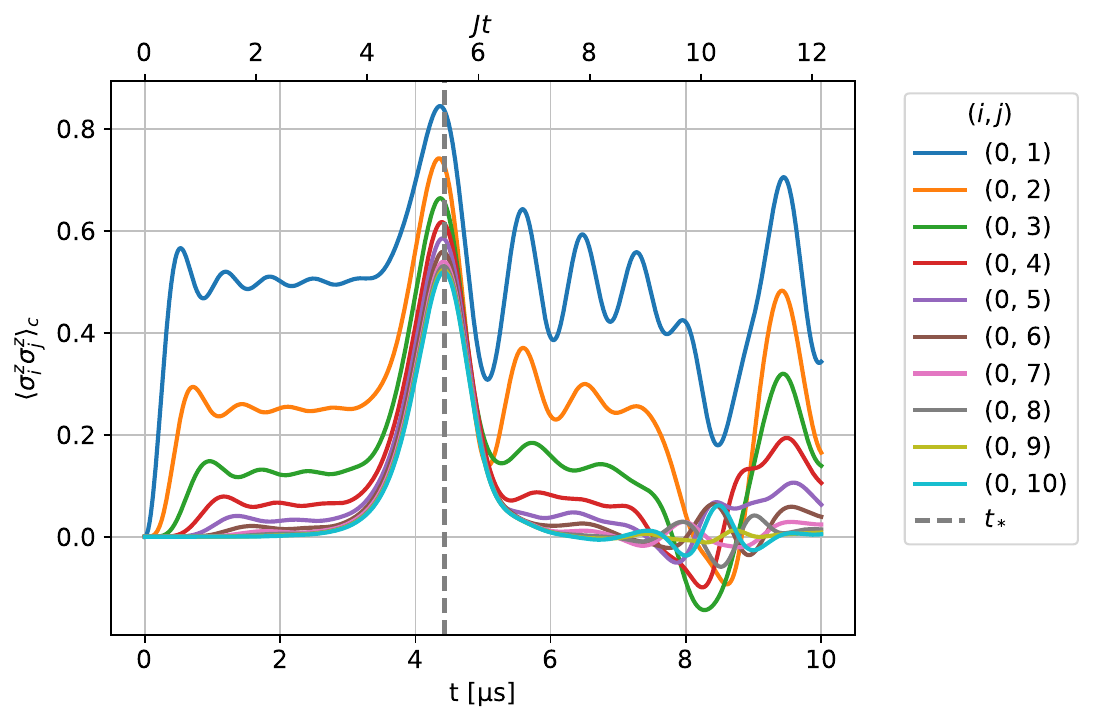}
		\caption{$\ket{+ \cdots +}$}
		\label{fig:2pt-corr-all-plus}
	\end{subfigure}
	
	\begin{subfigure}[c]{\linewidth}
		\centering
    	\includegraphics[width=\linewidth]{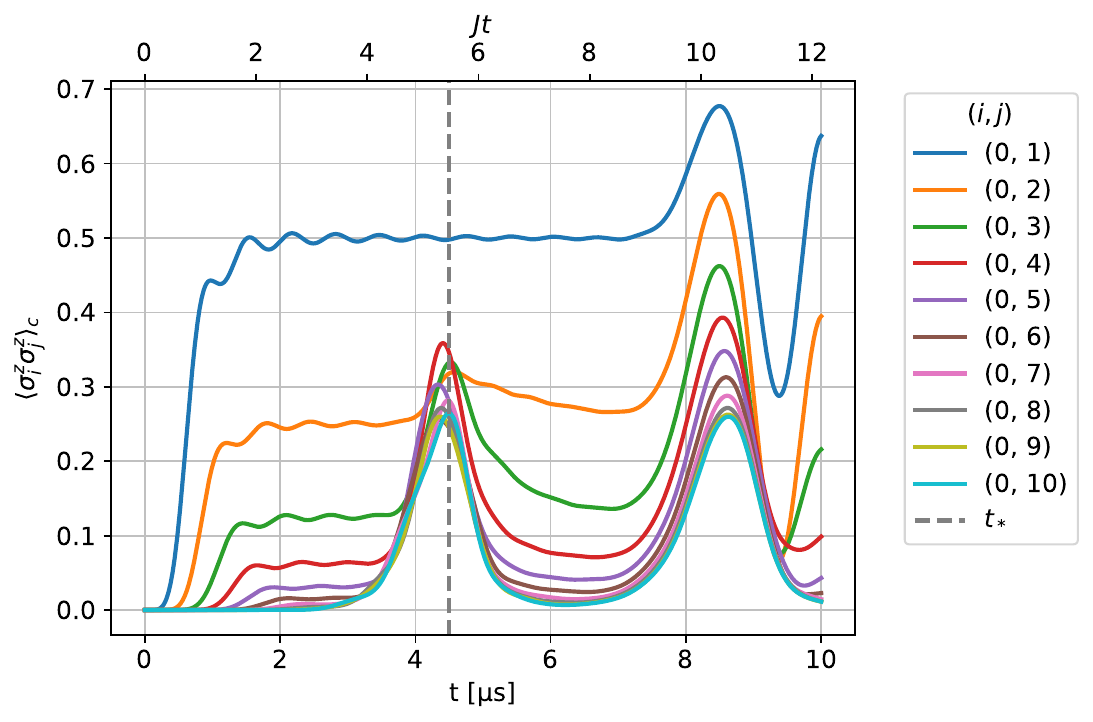}
		\caption{$\ket{\downarrow \cdots \downarrow}$}
	\end{subfigure}
	
	\begin{subfigure}[c]{\linewidth}
		\centering
	    \includegraphics[width=\linewidth]{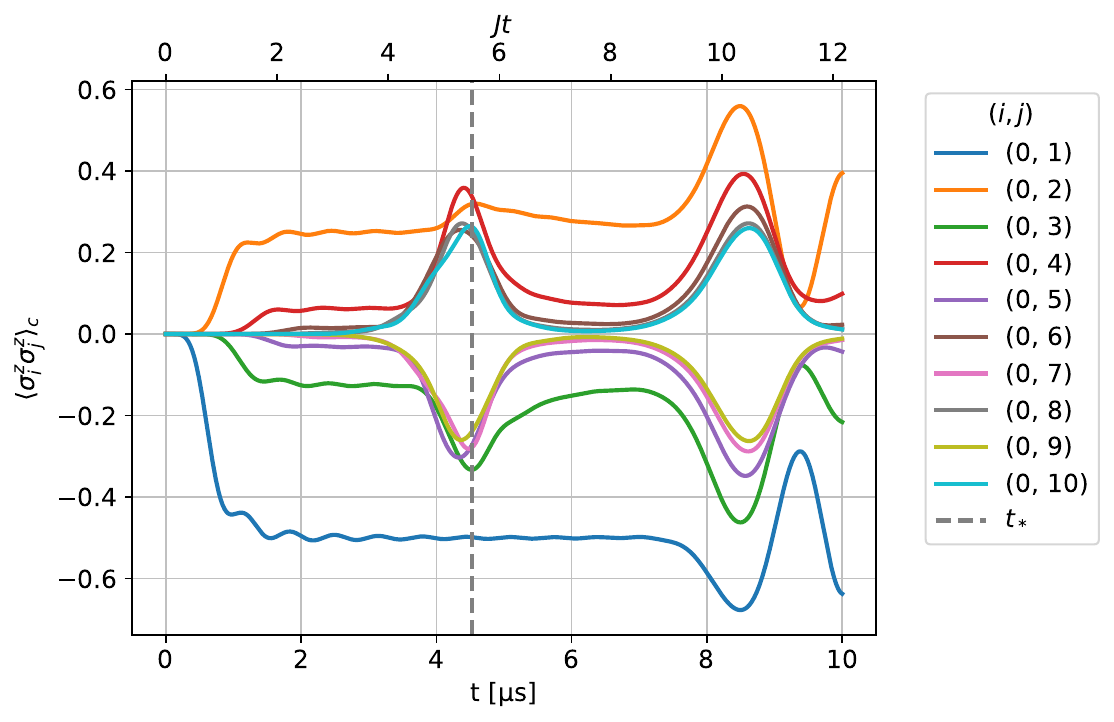}
		\caption{$\ket{\downarrow\uparrow \cdots \downarrow\uparrow}$}
	\end{subfigure}
	
	\caption{Correlation functions for $L = 20$, $g = 1$, $a = \qty{7.5}{\micro m}$ ($J \approx \qty{1.22}{rad \cdot \micro s^{-1}}$) for different initial states.
	}
	\label{fig:2pt-corr-all}
\end{figure}

In this appendix, we compare the correlation functions for all distances, showing that the overall dynamics studied in \cref{sec:analysis:peak} also occurs for other initial states (\cref{fig:2pt-corr-all}).

The correlation functions have a similar behavior: the more separated the sites are, the longer they stay zero.
This can be easily understood as a consequence of the Lieb-Robinson bound~\cite{LIEB_FiniteGroupVelocity_1972}.
Next, they start to increase and reach a plateau. This plateau corresponds to a time window where the state of the system is well described by a generalized Gibbs ensemble~\cite{rigolRelaxationCompletelyIntegrable2007}, with exponentially decaying correlations. 
Finally, when the last correlation function increases, they all peak together.
We see that the nearest-neighbor correlation functions for the down and AFM states immediately reach their peak value and have a plateau for a long time.

\section{Impact of the distance on the score}
\label[appsec]{app:impact_dist_score}

In this appendix, we study the behavior of the 2-point connected correlation functions
$g_d^{(2)}(t_*) := \braket{\sigma^z_i(t_*) \sigma^z_{i+d}(t_*)}_c$
at the surge time as a function of the distance $d$ between the sites, with site indices understood modulo $L$.

\begin{figure}
	\begin{subfigure}[c]{\linewidth}
		\centering
    	\includegraphics[width=\linewidth]{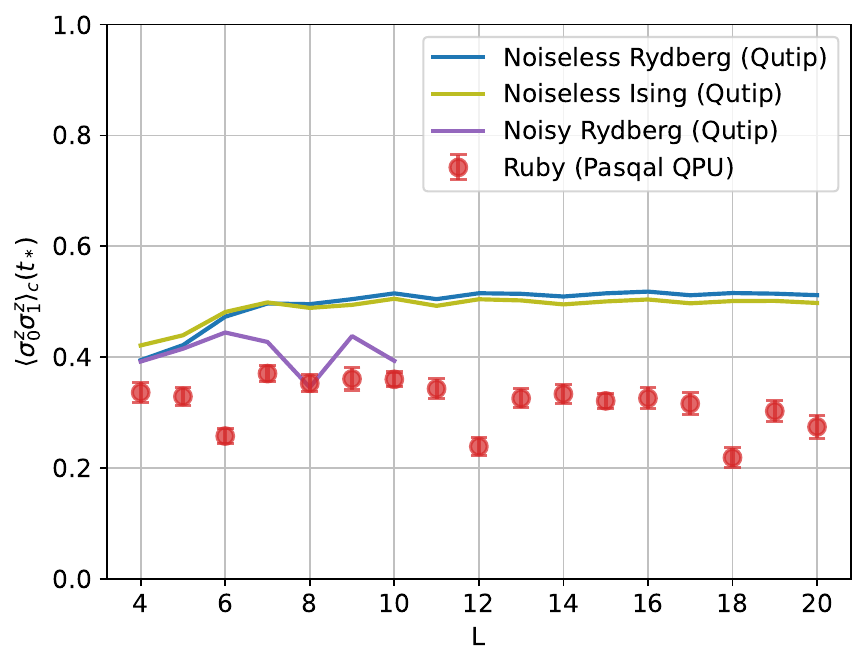}
		\caption{$d = 1$}
	\end{subfigure}
	
	\begin{subfigure}[c]{\linewidth}
		\centering
    	\includegraphics[width=\linewidth]{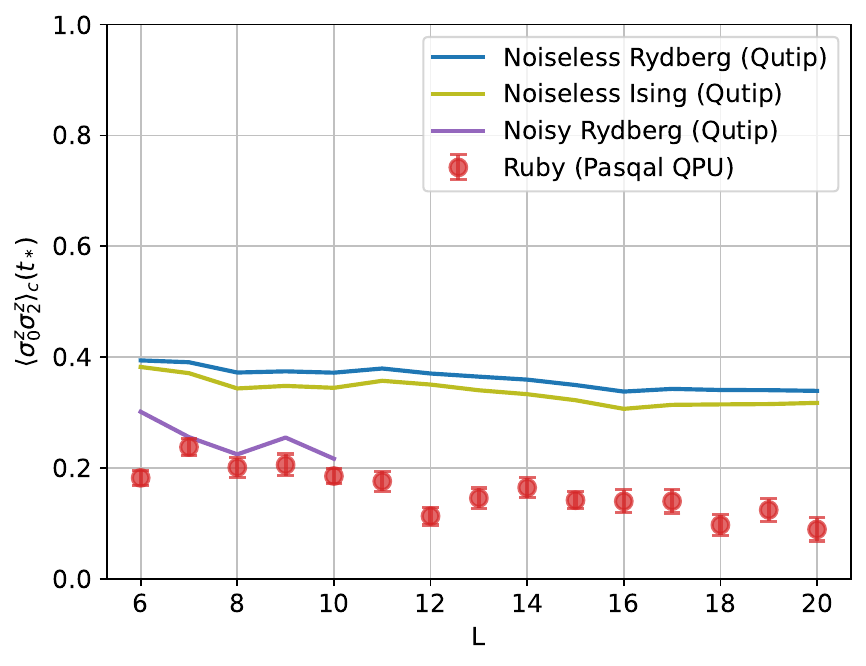}
		\caption{$d = 2$}
	\end{subfigure}
	
	\begin{subfigure}[c]{\linewidth}
		\centering
    	\includegraphics[width=\linewidth]{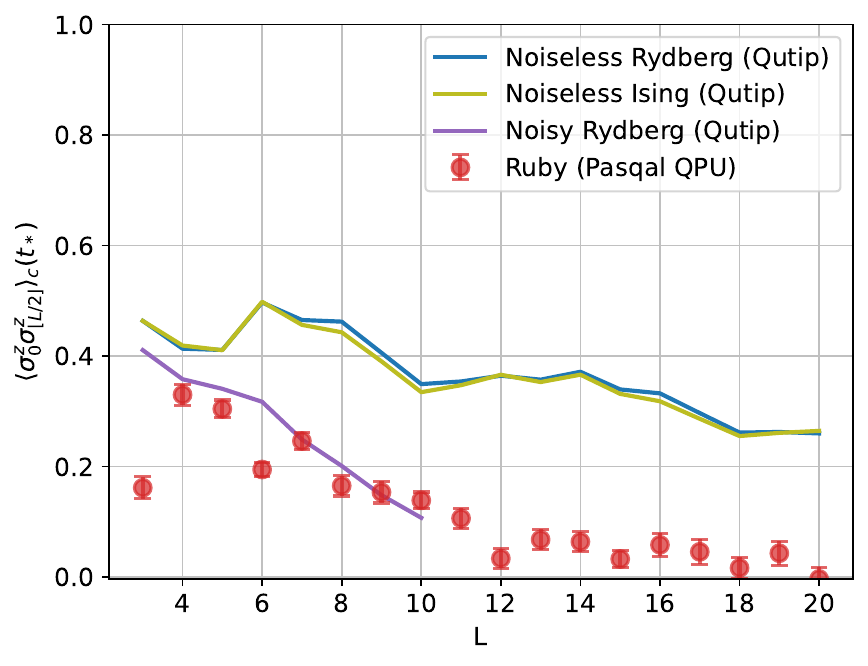}
			\caption{$d = \lfloor L/2 \rfloor$ (antipodal)}
	\end{subfigure}
	
	\caption{Connected correlation functions $g_d^{(2)}(t_*) = \braket{\sigma^z_i(t_*) \sigma^z_{i+d}(t_*)}_c$ at the surge time for $d = 1, 2, \lfloor L/2 \rfloor$.
	}
	\label{fig:2pt-corr-peak-distance}
\end{figure}

\begin{figure}
	\centering
	\includegraphics[width=\linewidth]{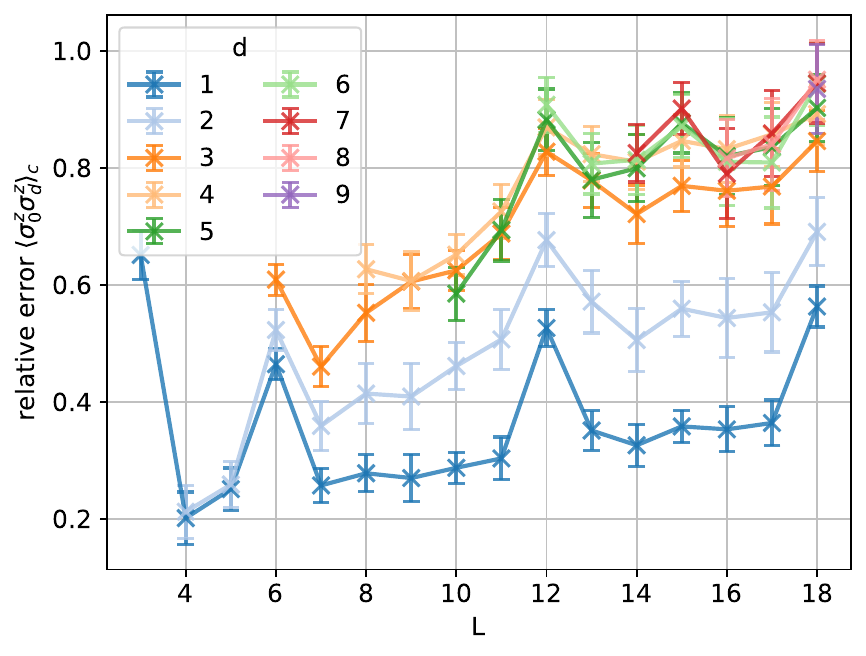}
	\caption{Relative error of $g_d^{(2)}(t_*)$ at the surge time as a function of $L$, for different values of $d$.
	}
	\label{fig:2pt-error-peak-distance}
\end{figure}

\Cref{fig:2pt-corr-peak-distance} compares the values of the correlation functions at the surge
time for different distances and for the noiseless Ising and Rydberg models, the noisy Rydberg simulation, and the experimental data.
First, we observe that the noise model accurately reproduces the experimental 2-point correlation data shown here.
Second, we see that the accuracy decreases at larger distances.
This can be seen more easily by plotting the relative errors as a function of the system size $L$, for different distances $d$, for the experimental data (\cref{fig:2pt-error-peak-distance}).
Hence, we conclude that errors at long distances dominate in $P_2(L)$, which means that the score is sensitive to the ability of the device to generate correlations at larger distances.
This is a feature of our benchmark: while it is easy to obtain strong correlations between neighboring qubits, obtaining a high score requires generating strong correlations at all distances.

\begin{figure}
	\centering
	\includegraphics[width=\linewidth]{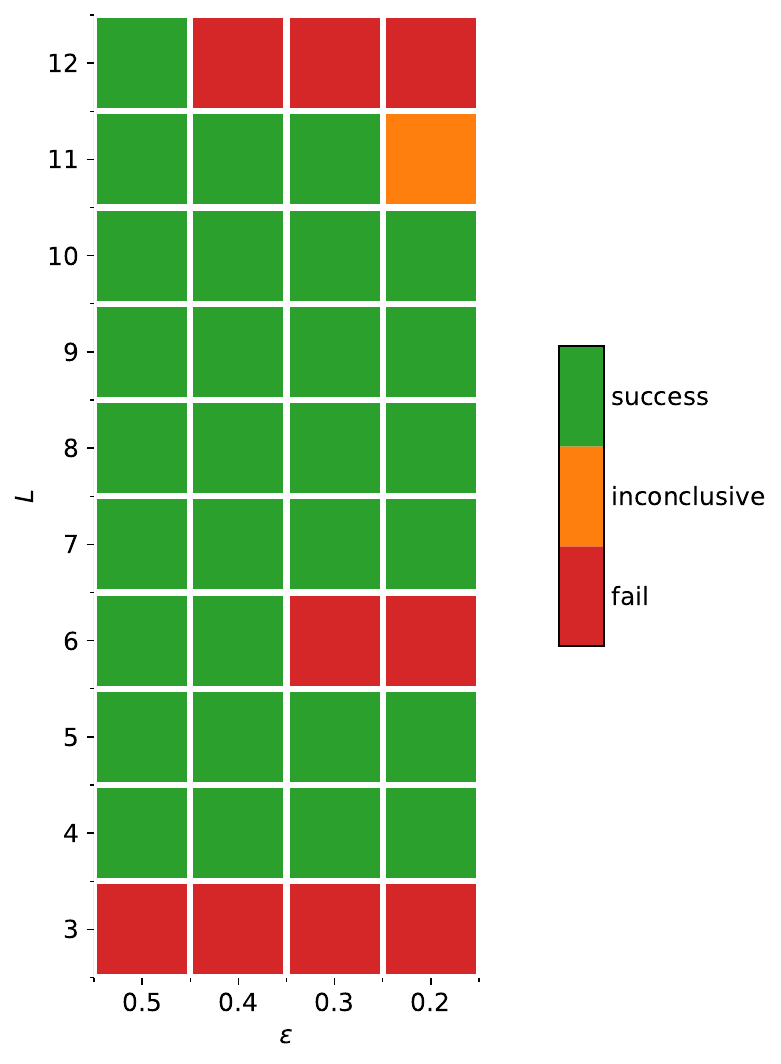}
	\caption{Volumetric plot obtained by considering only the relative error of $g_1^{(2)}(t_*) = \braket{\sigma^z_i(t_*) \sigma^z_{i+1}(t_*)}_c$.
	}
	\label{fig:volumetric-d1}
\end{figure}

This can be illustrated graphically by the volumetric plot (\cref{fig:volumetric-d1}) that would be obtained if the success test required the relative error of the 2-point correlation function at distance $d = 1$, $g_1^{(2)}(t_*)$, to lie below different thresholds.
In this case, we see that, with the same convention that the score starts at $L=4$, the mitigated values get a score of at least $S = 12$ for $\epsilon = 0.5$, and even $S = 12$ for $\epsilon = 0.3$ if the isolated $L = 6$ point is excluded.

\section{Analytical and numerical analysis of the surge time}
\label[appsec]{app:analytical_surge_time}

In this section, we use analytical and numerical methods to understand the behavior of correlations in the quenched Ising chain, both in a noiseless setting and in a noisy setting. In particular, we analyze the behavior of the surge time of the 2-point connected correlation function in a noiseless setting, and the behavior of the peak value in the presence of noise.

\subsection{Optimal measurement time}
\label{sec:analysis:peak}

This section aims to better understand and predict the surge time $t_*$ at which the connected correlation function reaches its maximal value (global correlation surge effect), in the case of the Ising Hamiltonian. The analysis is largely inspired by the work of Calabrese \emph{et al.} in~\cite{Calabrese_2012}.

\subsubsection{Notations and definitions}

We introduce the theoretical framework for describing the time evolution of an initial state $|\Psi_0\rangle = \ket{\downarrow \cdots \downarrow}$
under the TFIM Hamiltonian. Using the Jordan--Wigner transformation and diagonalization via Bogoliubov transformation in the
even parity sector
and in the odd parity sector,
we obtain an exact analytical expression for the state $|\Psi(t)\rangle$ in the form of BCS-type Gaussian states.

We consider the TFIM Hamiltonian with the parameter $g<1$:
\begin{equation}
	H_{\text{Ising}} =J \sum_{i=0}^{L-1}\sigma_i^{z}\sigma_{i+1}^z + gJ \sum_{i=0}^{L-1}\sigma_i^x
\end{equation}
where we identify $\sigma_L^z$ with $\sigma_0^z$ (periodic boundary conditions).

The spin operators are mapped onto fermionic ones using the standard Jordan--Wigner transformation:
\begin{equation}
\begin{aligned}
    \sigma_j^x& = 1-2n_j, \\
    \sigma_j^y& = \I \left ( \prod_{j'<j} \e^{\I\pi n_{j'}} \right ) (c_j - c_j^\dagger), \\
    \sigma_j^z& = \left ( \prod_{j'<j} \e^{\I\pi n_{j'}} \right) (c_j + c_j^\dagger),\\
\end{aligned}
\end{equation}
where $n_j = c_j^\dagger c_j$ is the number operator.
The resulting Hamiltonian is: 
\begin{equation}
\begin{aligned}
    H_{\text{Ising}}
        &
        =  J \sum_{i=0}^{L-2}[c_i^\dagger c_{i+1} + c_i^\dagger c_{i+1}^\dagger + \text{h.c.}] 
            \\ & \quad
            + gJ \sum_{i=0}^{L-1}[1-2c_i^\dagger c_i] 
            \\ & \quad
            + \e^{\I\pi N_f} J (c_{L-1}^\dagger c_0 + c_{L-1}^\dagger c_0^\dagger + \text{h.c.}).
\end{aligned}
\end{equation}
where $N_f = \sum_i c_i^\dagger c_i$ is the total number of fermions.

This is not directly a quadratic form due to the factor $e^{\I\pi N_f}$. But the fact that $H$ conserves the fermion parity, $[e^{\I\pi N_f},H_{\text{Ising}}] = 0$, leads us to consider the eigenspace decomposition of the operator $e^{\I\pi N_f}$.
The eigenspace of eigenvalue $1$ (resp.~$-1$) is called the Neveu-Schwarz (resp.~Ramond) sector and corresponds to the states having an even (resp.~odd) number of fermions.
In each of these sectors, $H_{\text{Ising}}$ is effectively quadratic:
\begin{equation}
H_{\text{Ising}} = P^\mathrm{R}H^\mathrm{R}P^\mathrm{R} + P^\mathrm{NS}H^\mathrm{NS}P^\mathrm{NS}
\end{equation}
with: 
\begin{equation}
\begin{aligned}
    H^a
        &
        =  J \sum_{i=0}^{L-2}[c_i^\dagger c_{i+1} + c_i^\dagger c_{i+1}^\dagger + \text{h.c.}] 
            \\ & \quad
            + g J \sum_{i=0}^{L-1}[1-2c_i^\dagger c_i] 
            \\ & \quad
            +\epsilon_a J (c_{L-1}^\dagger c_0 + c_{L-1}^\dagger c_0^\dagger + \text{h.c.})
\end{aligned}
\end{equation}
and
\begin{equation}
	P^a = \frac{1 - \epsilon_a\, \e^{\I\pi N_f}}{2}
\end{equation}
where $a = \mathrm{R}, \mathrm{NS}$, $\epsilon_\mathrm{R} = 1$, $\epsilon_\mathrm{NS} = -1$. $H^\mathrm{R}$ corresponds to periodic boundary conditions for the fermions, and $H^\mathrm{NS}$ corresponds to antiperiodic boundary conditions.

As quadratic fermionic Hamiltonians, $H^\mathrm{R}$ and $H^\mathrm{NS}$ can both be diagonalized with a Bogoliubov transformation after the appropriate Fourier transformation of the modes, namely with different lattice momenta $\mathcal{K}_a$ for the two sectors due to different boundary conditions:
\begin{equation}
    \begin{aligned}
	\mathcal{K}_\mathrm{R}
	    &
	    = \left\{\frac {2n\pi}{L}, ~n = -\frac L 2, ..., \frac L 2 - 1 \right\},
	   \\
	\mathcal{K}_\mathrm{NS}
	    &
	    = \left\{\frac {2(n+1/2)\pi}{L}, ~n = -\frac L 2, ..., \frac L 2 - 1 \right\}.
    \end{aligned} 
\end{equation}

We define, for $k \in \mathcal{K}_a$, the Fourier-transformed fermionic modes: 
\begin{equation}
c_k = \frac{1}{\sqrt{L}}\sum_{j=0}^{L-1} \e^{\I kj}c_j,
\end{equation}
and the Bogoliubov fermions $\gamma_k$ (for $k\ge0$): 
\begin{equation}
\begin{aligned}
    c_k
        &
        = \cos (\theta_k / 2)\gamma_k+ \I \sin(\theta_k / 2)\gamma_{-k}^\dagger,
    \\
    c_{-k}^\dagger
        &
        = \I \sin(\theta_k / 2)\gamma_k+ \cos (\theta_k / 2)\gamma_{-k}^\dagger,
\end{aligned}
\end{equation}
with the Bogoliubov angle $\theta_k$ defined by: 
\begin{equation}
\e^{\I \theta_k} = \frac{g-\e^{\I k}}{\sqrt{1+g^2-2g\cos k}}.
\end{equation}

With these definitions, the Hamiltonians $H^\mathrm{R}$ and $H^\mathrm{NS}$ are effectively diagonalized: 
\begin{equation}
	H^a = \sum_{k\in \mathcal K_a} \varepsilon_k \left(\gamma_k^\dagger \gamma_k - \frac{1}{2} \right)
\end{equation}
with the (negative) eigenenergies
\begin{equation}
	\varepsilon_k = -2J\sqrt{1 + g^2 - 2g~\cos k},
\end{equation}
and state of maximal energy $E_\mathrm{max}^a$ (in our conventions) $|\emptyset,g\rangle^a$.
We note that since for $g<1$, $E_\mathrm{max}^\mathrm{R} - E_\mathrm{max}^\mathrm{NS} = \mathcal{O}(\e^{-L})$, we consider these two energies to be equal. This exponential decay is observed, for example, in~\cite[page 19]{Mbeng_2024}.

\begin{figure}
    \includegraphics[width=\linewidth]{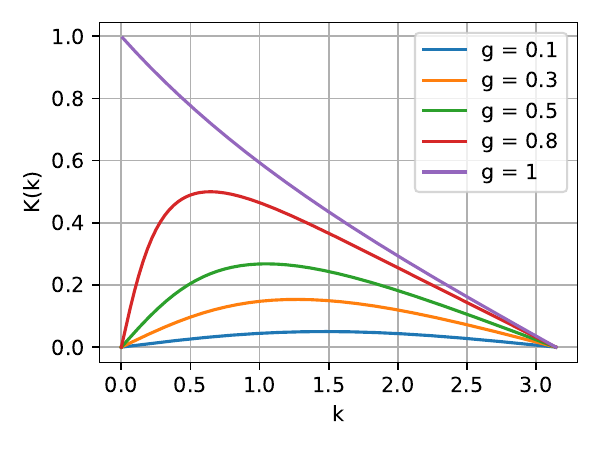}
    \caption{Function $K(k)$ for different values of $g$.}
    \label{fig:K_func}
\end{figure}

Following the calculations of~\cite{Calabrese_2012}, we can derive an exact expression of $|\Psi(t)\rangle$:
\begin{equation}
\label{eq:state-RNS}
	|\Psi(t)\rangle = \frac{1}{\sqrt{2}} \sum_{a= \mathrm{R},\mathrm{NS}} \frac{|\Phi(t)\rangle^a}{\sqrt{^a\langle \Phi(t)|\Phi(t)\rangle^a}},
\end{equation}
with:
\begin{equation}
    \label{eq:Phi_t}
    \begin{split}
    |\Phi(t)\rangle^a
        &
        = \e^{-\I E_\mathrm{max}^at} 
    \\ &
    \times \exp\left[ \I \!\! \sum_{0<k\in \mathcal K_a} \!\! K(k) \e^{-2\I \varepsilon_k t} \gamma_{-k}^\dagger \gamma_k^\dagger\right] |\emptyset,g\rangle^a,
    \end{split}
\end{equation}
and the function $K(k)$ (plotted in \cref{fig:K_func}):
\begin{equation}
K(k) = \frac{g \sin k}{1 - g\cos k +  \sqrt{1 + g^2 - 2g\cos k} }.
\end{equation}
The states $|\Phi(t)\rangle^a$ are Gaussian states, in a BCS form. $K(k)$ is related to the number of excitations in mode $k$ at $t=0$:
\begin{equation}
	{}^a\langle \Phi(t=0) |\gamma_k^\dagger \gamma_k|\Phi(t=0)\rangle^a = \frac{K(k)^2}{1 + K(k)^2}.
\end{equation}
For $g=0$ we have $K(k) = 0$, and it increases with $g$. But it remains smaller than $1$ as long as $g<1$.

\begin{figure}
    \includegraphics[width=\linewidth]{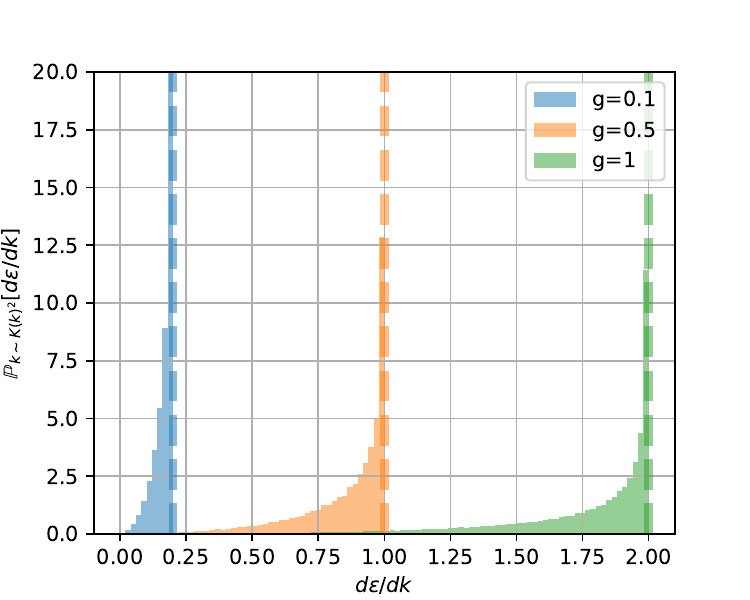}
    \caption{Distribution of weighted quasiparticle velocities $\dd \varepsilon(k) / \dd k$, for different values of $g$. Most velocities are close to $v_\mathrm{max}$ (vertical dashed lines).}
    \label{fig:velocity_distribution}
\end{figure}

\subsubsection{Lehmann representation of the 2-point correlation functions}

In this section, we derive an expression of the (disconnected)
correlation function $\langle \Psi(t)|\sigma_i^z \sigma_{i+\ell}^z |\Psi(t)\rangle$ for large $L$ as an expansion in powers of $K(k)$. The analysis of the first term of the expansion yields a condition on the time at which the correlation function reaches its maximal value. In practice, this time is very close to the time of maximum for the connected correlation function, which justifies our approach. 

Since $\sigma_i^z \sigma_{i+\ell}^z$ does not change the fermion parity,
cross terms between both sectors vanish, and 
we can express the correlation function $\langle \sigma_i^z \sigma_{i+\ell}^z \rangle$ as:
\begin{equation}
\begin{aligned}
	&
	\langle \Psi(t)|\sigma_i^z \sigma_{i+\ell}^z |\Psi(t)\rangle
	\\ & \;
	= \frac{1}{2} \sum_{a = \mathrm{R},\mathrm{NS}} \frac{{}^a\langle \Phi(t)|\sigma_i^z \sigma_{i+\ell}^z|\Phi(t)\rangle^a}{{}^a\langle \Phi(t)|\Phi(t)\rangle^a}.
\end{aligned}
\end{equation}
Let us expand the correlation function in the R sector ${}^\mathrm{R}\langle \Phi(t)|\sigma_{i}^{z} \sigma_{i+\ell}^{z}|\Phi(t)\rangle^\mathrm{R}$ according to Eq.~\eqref{eq:Phi_t} (the analysis in the NS sector is similar):
\begin{equation}
\begin{aligned}
    & {}^\mathrm{R}\langle \Phi(t)|\sigma_{i}^{z} \sigma_{i+\ell}^{z}|\Phi(t)\rangle^\mathrm{R}
        \\ &
        = \sum_{l,n,r=0}^{\infty} \frac{{\I}^{n-l}}{n! \, l!} \sum_{\substack{0<k_1\neq \cdots\neq k_n \in \mathcal{K}_\mathrm{R} \\ 0<p_1\neq \cdots\neq p_l \in \mathcal{K}_\mathrm{R}}} \sum_{q_1, \ldots, q_r \in \mathcal{K}_\mathrm{NS}} 
            \\ & \quad
            \times \frac{1}{r!} \left[\prod_{j=1}^{n} K(k_j)\right] \left[\prod_{j=1}^{l} K(p_j)\right]
            \\ & \quad
            \times \e^{-2\I t\left[\sum_{j=1}^l \varepsilon_{p_j} - \sum_{j=1}^n \varepsilon_{k_j}\right]}
            \\ & \quad
            \times {}^\mathrm{R}\langle -k_1, k_1, \ldots, -k_n, k_n|\sigma_{i+\ell}^{z}|q_1, \ldots, q_r\rangle_\mathrm{NS}
            \\ & \quad
            \times {}^\mathrm{NS}\langle q_r, \ldots, q_1|\sigma_{i}^{z}|p_1, -p_1, \ldots, p_l, -p_l\rangle^\mathrm{R}
            \\ &
        = \sum_{l,n,r=0}^{\infty} D_{(2n|r|2l)}.
\end{aligned}
\end{equation}
Each term of the sum oscillates in time with a frequency $\sum_{j=1}^l \varepsilon_{p_j} - \sum_{j=1}^n \varepsilon_{k_j}$. When $n\ne l$, these frequencies are broadly distributed across the spectrum, resulting in destructive interference and a vanishing contribution to the correlation function as soon as $t>0$.

On the other hand, when $n=l$, the spectrum will be peaked around particular values that are multiples of $4v_\mathrm{max} \pi / L$,
where $v_\mathrm{max}$ is the maximal group velocity (Lieb--Robinson velocity).
It can be obtained analytically for $g<1$ as:
\begin{equation}
\label{eq:max-velocity}
v_\mathrm{max} = \Bigg\lVert\frac{\partial\varepsilon_k}{\partial k}\Bigg\rVert_\infty = 2gJ.
\end{equation}
Indeed, when the values taken by the  $\{\varepsilon_{k_j}\}$ are close to the $\{\varepsilon_{p_j}\}$, each pair has a contribution 
\begin{equation*}
   \varepsilon_p-\varepsilon_k \approx (p-k)\frac{\dd \varepsilon}{\dd k}.
   \end{equation*}
In turn, Fig. \ref{fig:velocity_distribution} shows that $\frac{\dd \varepsilon}{\dd k}$ is close to $v_\mathrm{max}$, leading to:
\begin{equation*}
   (p-k)\frac{\dd \varepsilon}{\dd k} \approx \frac{4m\pi v_\mathrm{max}}{L},\; m\in \mathbb{Z}.
\end{equation*}

With this simple observation, we see that the correlation function has a pseudo period of
\begin{equation}
t_\mathrm{F} = L/(2v_\mathrm{max}). \label{eq:t_F_def}
\end{equation} 

Among the remaining $D_{(2n|r|2n)}$ terms, only the $D_{(2n|2n|2n)}$ ones are actually predominant due to the factor $ {}^\mathrm{NS}\langle q_r, \ldots, q_1|\sigma_{i}^{z}|p_1, -p_1, \ldots, p_l, -p_l\rangle^\mathrm{R}$.
Last, if $g$ (and $K(k)$) is sufficiently small, the first term $D_{(2|2|2)}$ will govern the dynamics:
\begin{equation}
    \begin{aligned}
    &
    D_{(2|2|2)}
    \\ &
    = \frac{1}{2} \sum_{\substack{0 < k\ne p \in \mathcal{K}_\mathrm{R} \\ q_1,q_2\in \mathcal{K}_\mathrm{NS}}} \!\!
        K(k) K(p)e^{-2\I t(\varepsilon_p - \varepsilon_k)}\e^{\I \ell(q_1 + q_2)}
    \\ &  \;
    \times {}^\mathrm{R}\langle -k,k|\sigma_i^z|q_1,q_2\rangle^\mathrm{NS} \; {}^\mathrm{NS}\langle q_1,q_2|\sigma_{i+\ell}^z|-p,p\rangle^\mathrm{R}
    \\ &
    \approx \frac{2(1-g^2)^{1/4}}{L^2}
        \\ & \;
        \times \!\!\!\! \sum_{k\ne p\in \mathcal{K}_\mathrm{R}} \!\!\!
            \e^{-2\I t(\varepsilon_k - \varepsilon_p)} \frac{(\varepsilon_k + \varepsilon_p)^2}{\varepsilon_k^2\varepsilon_p^2}
            \frac{K(k)K(p)}{(\cos(k)- \cos(p))^2}
        \\ & \quad
        \times \left[
            \sin(k)\sin(p)\left(\frac{\varepsilon_{k} + \varepsilon_p}{2}\right)^2\right.
            \\ & \qquad
            + \varepsilon_k\varepsilon_p\sin^2\left(\frac{k-p}{2}\right)\cos\big(\ell(k+p)\big)
            \\ & \qquad
            - \varepsilon_k\varepsilon_p\left.\sin^2\left(\frac{k+p}{2}\right) \cos\big(\ell(k-p) \big)
        \right].
    \end{aligned}
\end{equation}
The last approximation corresponds to Eq. (179) of~\cite{Calabrese_2012}.

\begin{figure}[h]
    \includegraphics[width=\linewidth]{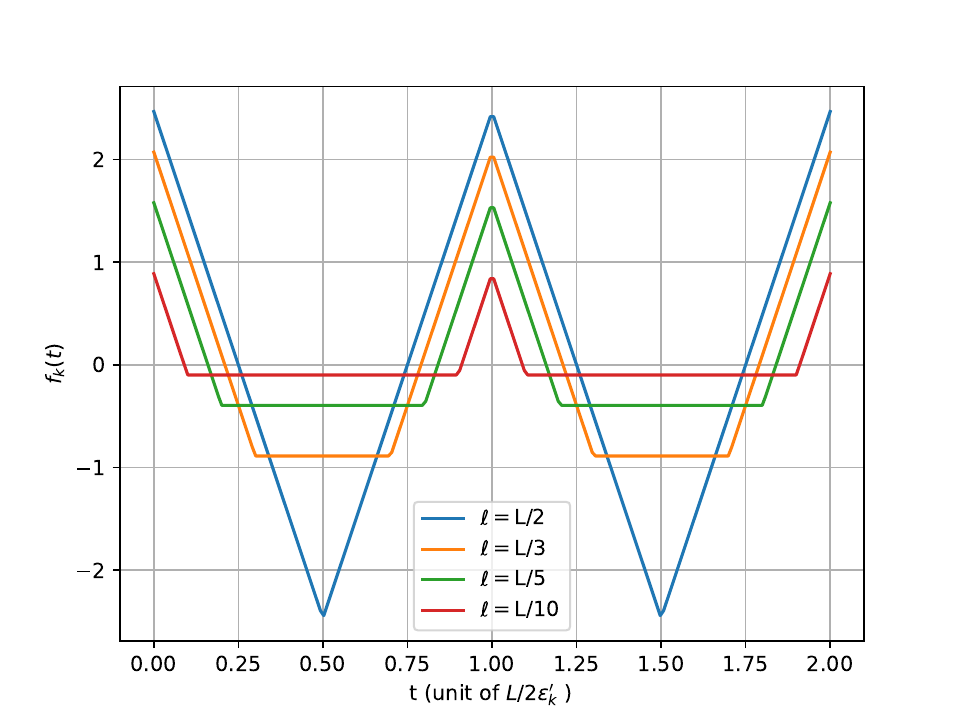}
    \caption{Functions $f_{k,\ell}(t)$ for different values of $\ell / L$.}
    \label{fig:f_plot}
\end{figure}

We  now assume that $k$ and $p$ are very close in the preceding expression and expand in their difference $k-p = \frac{2m\pi}{L}$.
Due to the factor $[\cos(k) - \cos(p)]^{-2}$ that only takes large values for $k\approx p$, this approximation provides a result that closely matches the exact value. We arrive at :
\begin{equation}
    \text{Re} [D_{(2|2|2)}] \propto \sum_{0 < k \in \mathcal{K}_\mathrm{R}} K(k)^2 f_{k,\ell}(t)
\end{equation}
where we defined
\begin{equation}
\begin{aligned}
    f_{k,\ell}(t)
        &
        = \sum_{m\in \mathbb{N}^*} \frac{1}{m^2}\cos \left(\frac{4 \pi tm\varepsilon_k'}{L}\right)
         \\ & \qquad
         \times\left(1 - \cos \left(\frac{2m\ell \pi}{L}\right)\right), 
\end{aligned}
\end{equation}
with the shorthand $\varepsilon_k' = \dd \varepsilon/\dd k$.
The functions $f_{k,\ell}(t)$, shown in Fig. \ref{fig:f_plot}, have a period of $\frac L {2\varepsilon_k'}$.
We even have a closed form of this function (straightforward calculations with Spence's function).
For $0\le t \le \frac L {2\varepsilon_k'}$:
\begin{equation}
\begin{aligned}
    &f_{k,\ell}(t)
    \\ &
    =
    \begin{cases}
    \displaystyle
    f_{k,\ell}(0) - t \frac {2\pi^2\varepsilon_k'}{L} & 0 \le t \le t_1
    \\
    \displaystyle
    -\frac {\ell^2 \pi^2}{L^2} & t_1 \le t \le t_2
    \\
    \displaystyle
    f_{k,\ell}(0) - \left(\frac{L}{2\varepsilon_k'} - t \right) \frac {2\pi^2\varepsilon_k'}{L}&  t_2 \le t \le \frac L {2\varepsilon_k'}
    \end{cases} 
\end{aligned}
\end{equation}
with
\begin{equation}
    \begin{gathered}
    t_1 = \frac {\ell}{2\varepsilon_k'},
    \qquad
    t_2 = \frac L {2\varepsilon_k'} - \frac {\ell}{2\varepsilon_k'},
    \\
    f_{k,\ell}(0) = \pi^2 \frac \ell L \left(1 - \frac \ell L\right).
    \end{gathered}
\end{equation}

Let us summarize our results. We have found a simplified expression for the correlation function: 
\begin{equation}
    \langle \sigma_i^z \sigma_{i+\ell}^z (t) \rangle \approx A + B \sum_{0<k\in \mathcal{K}_R} K(k)^2f_{k,\ell}(t),
\end{equation}
where $A$ and $B$ are two unknown constants. We actually do not  need to compute them since our aim is only to find the time of the first maximum. 

Each contribution $f_{k,\ell}(t)$ is maximal at $t= \frac L {2\varepsilon_k'}$. Looking at \cref{fig:velocity_distribution}, we see that by summing all these contributions, the connected correlation function reaches its maximum at a time $t_* \gtrsim t_\mathrm{F}$.

\begin{figure}
\includegraphics[width=\linewidth]{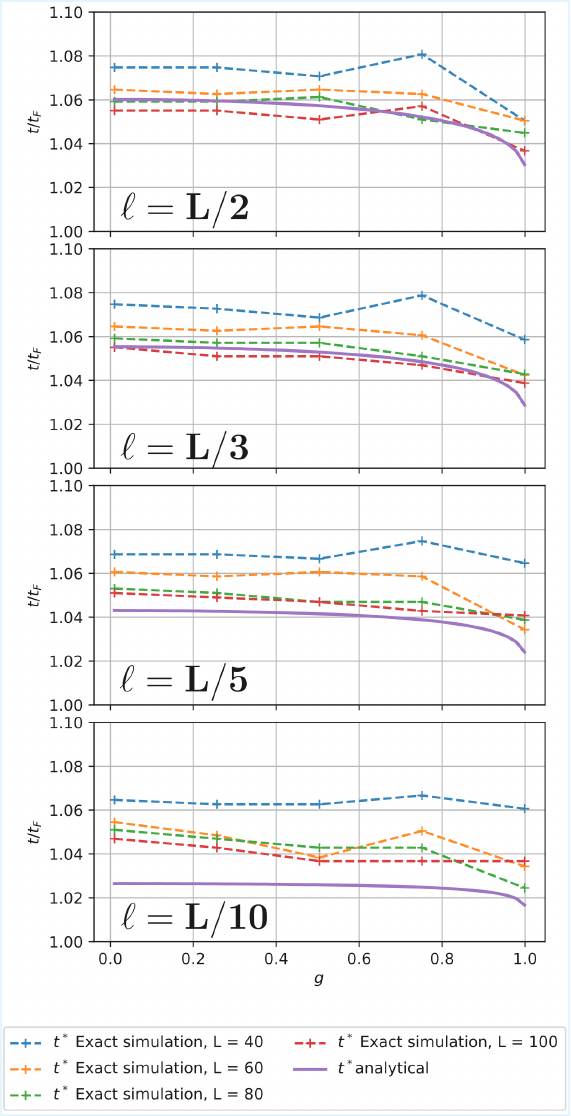}
\caption{Time $t_*$ at which the correlation function reaches its maximum, as a function of $g$, for different values of $\ell/L$. $t_*$ is computed numerically in exact simulations for different values of $L$ (dashed lines), and with Eq.~\eqref{eq:t_star_condition} (purple solid line). Times are rescaled by $t_\mathrm{F} = L/2v_\mathrm{max}$ in order to remove dependence on $L$.}
\label{fig:t_star_estimate}
\end{figure}

Let us evaluate $t_*$ more quantitatively. 
By taking the derivative and approximating the discrete sum by an integral, we have a condition on the time of maximum: 
\begin{equation}
    \int_0^\pi K(k)^2 \frac{\dd f_{k,\ell}}{\dd t}\Bigg|_{t=t_*} \dd k = 0.
    \label{eq:t_star_condition}
\end{equation}
Since $f_{k,\ell}$ is a function of $t/L$ only, the solution $t_*$ of this equation can be seen to be directly proportional to $L$. 
This again reflects the ballistic nature of the propagation and is in agreement with direct numerical calculations (see Eq. \eqref{eq:t_star_num} above).

Let us solve this equation numerically. We obtain the purple curve shown in \cref{fig:t_star_estimate}. 
We see that $t_*$ varies very little as a function of $\ell$ (and $g$).
Last, let us compare this analytical estimate with the exact maximum obtained by a full free fermion simulation. 
We observe that the exact $t_*$ (dashed lines) is very close to our analytical estimate (purple solid line).
This provides an explanation for the global correlation surge phenomenon, where correlations at all distances peak at (almost) the same time.

\subsection{Impact of dephasing noise}

A question that naturally arises is the impact of noise on the MBQS. In this section, we investigate numerically a simple noise model in which we only take into account the dephasing noise. This allows us to find an empirical quantitative relation between the final score and the noise amplitude $\gamma$.

We assume that the system dynamics is governed by the Lindblad master equation:
\begin{equation}
\partial_t \rho = -\I[H,\rho] +\sum_{m=0}^{L-1} \left(L_m\rho L_m^\dagger - \frac{1}{2} \{L_m^\dagger L_m,\rho\}\right)
\end{equation}
where the jump operators are $L_m = \sqrt{\gamma} \sigma_m^z$, with $\gamma$ the dephasing rate.

We define the following ratio
\begin{equation}
    \eta (\gamma) =\frac{\text{max}_t[g^{(2)}_{L/2} (t,\gamma)]}{\text{max}_t[g^{(2)}_{L/2} (t,\gamma= 0)]}
\end{equation}
between the noisy and the noiseless antipodal correlations (at the time where they are maximal).

\begin{figure}
    \includegraphics[width=\columnwidth]{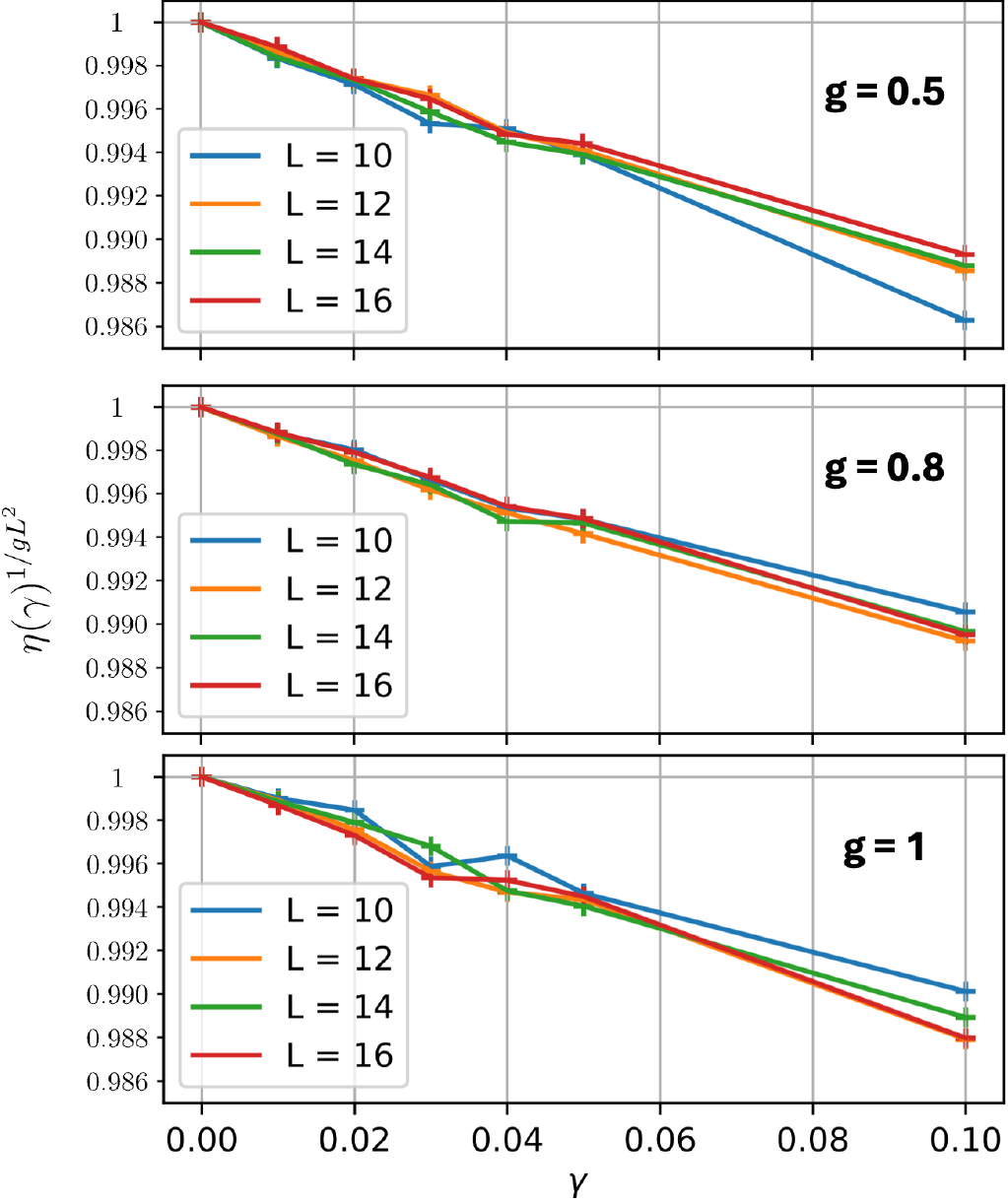}
    \caption{$\eta(\gamma)^{1/(gL^2)}$ as a function of the noise amplitude $\gamma$, for different values of $g$ and $L$. Simulations are executed using Eviden's Qaptiva software (with MPSTraj backend).
    }
    \label{fig:decrease_rho_gamma}
\end{figure}

Quantitatively, solving the Lindblad equation numerically, we find, in \cref{fig:decrease_rho_gamma}, a decay that is well described by the following relation:
\begin{equation}
\label{eta_empirical}
\eta(\gamma) = \e^{-\beta \gamma g L^2},
\end{equation}
where $\beta \approx 0.12$ is a fitting constant.
In other words, dephasing noise suppresses correlations, as expected. 
The exponential decrease in $L^2$ can be rationalized with a simple counting argument: there are $L$ jump operators acting over an effective timescale $t_*$ that is itself proportional to $L$.  

Let us now conclude with an estimate of the MBQS in the presence of noise.
Our previous definition of the score is based on the average of 2-point correlations at all distances.
If only the antipodal correlation is considered, we have a first-order approximation of the score that a quantum computer subject to dephasing noise $\gamma$ would achieve.

With Eq.~\eqref{eta_empirical}, the system sizes $L$ that have a relative error $[1 - \eta(\gamma)]$ below the threshold $\varepsilon$ are characterized by :
\begin{equation}
     \e^{-\beta \gamma g L^2} \ge 1-\varepsilon
\end{equation}

The limiting size for $g=1$---which defines the MBQS---is therefore : 
\begin{equation}
S = \sqrt{\frac{-\log(1 - \varepsilon)}{\beta \gamma}} \propto \gamma^{-1/2}
\end{equation}
Substituting the values $\gamma = 1/20$ and $\varepsilon = 0.5$, we find $S \approx 11$, which is of the same order of magnitude as the experimental values, although other sources of noise are dominant (\cref{app:noise}).\footnote{On early devices, $\gamma \simeq 0.2$, which implies $S \approx 5$.}

\section{Matrix-product state simulations -- benchmark score of a classical computer}
\label[appsec]{app:MPS}

\begin{figure}
    \centering
    \includegraphics[width=\linewidth]{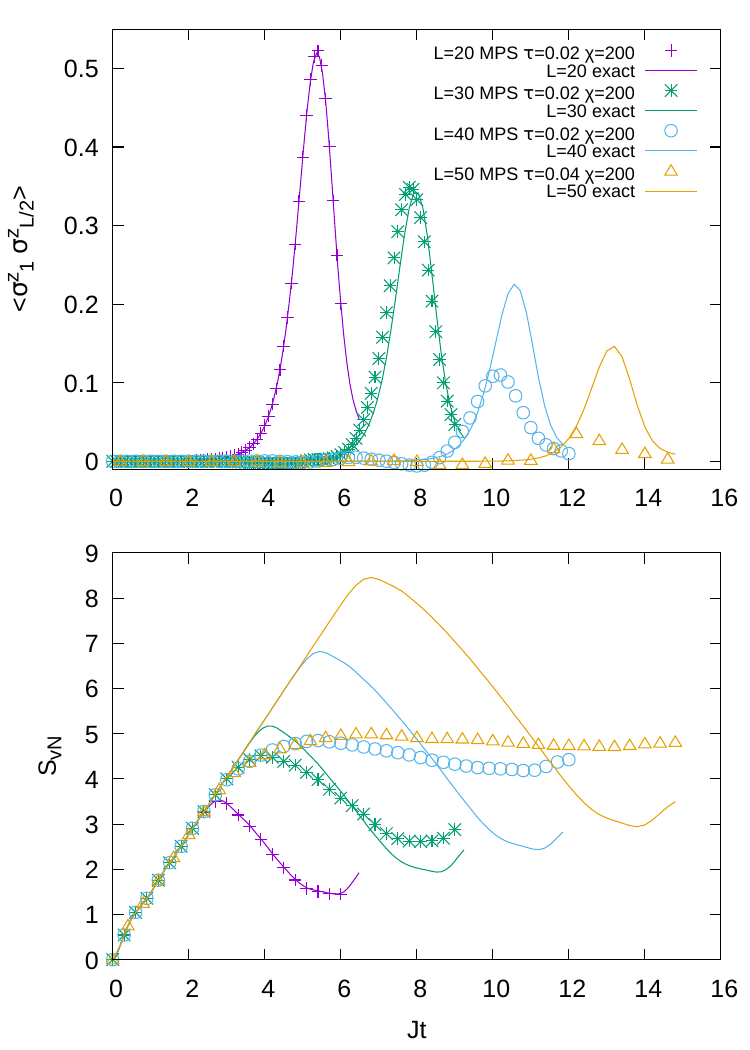}
    \caption{MPS simulations with maximum bond dimension $\chi=200$. Upper panel: antipodal correlation $\langle \sigma^z_0 \sigma^z_{\lfloor L/2 \rfloor}\rangle$ as a function of rescaled time $Jt$ for $L=20, 30, 40$ and $50$.
    Simulations with this maximum bond dimension resolve the correlation peak for $L=30$ (green) but not for $L=40$ (blue). Bottom panel: von Neumann entanglement entropy for a central bipartition of the system in two equal halves.}
    \label{fig:mps200}
\end{figure}

\begin{figure}
    \centering
    \includegraphics[width=\linewidth]{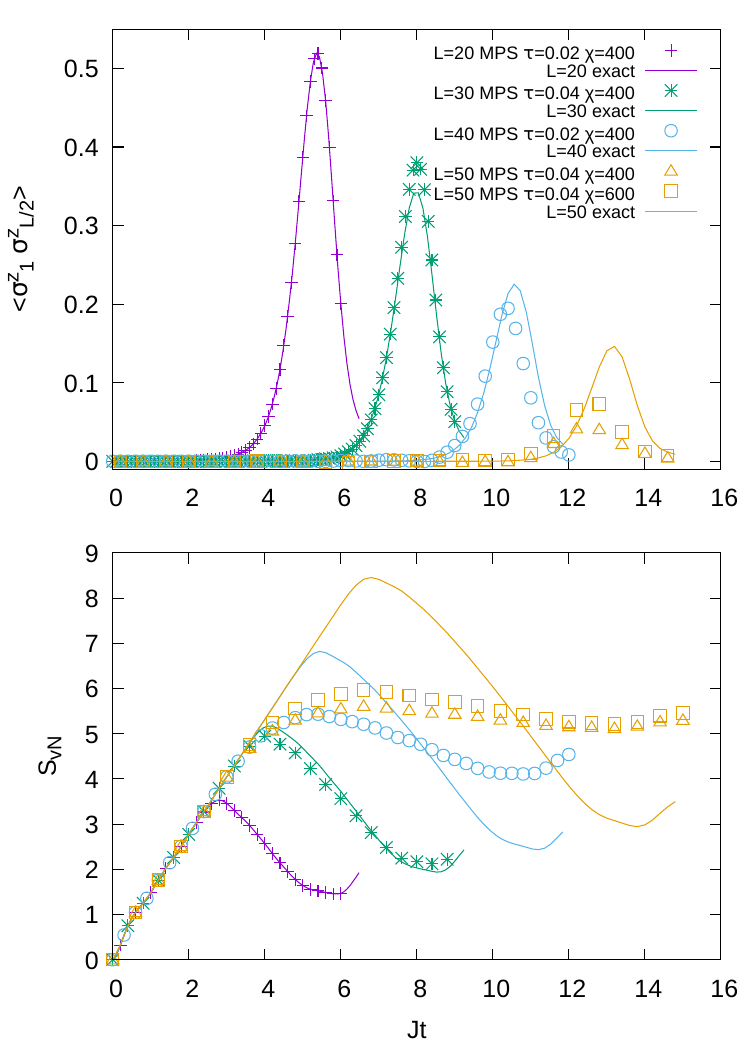}
    \caption{Same as \cref{fig:mps200} with maximum bond dimensions $\chi=400$ and $\chi=600$. Simulations with $\chi=400$ resolve the correlation peak for $L=40$ (blue) but not for $L=50$ (yellow).
    For $L=50$ some data with $\chi=600$ are also displayed (square symbols). In that case the numerical result for the peak height is still only $\sim50\%$ of its exact value.}
    \label{fig:mps400}
\end{figure}

In this section we show to what extent the present quench problem can be simulated with MPS on a classical computer. The initial state is chosen here to be $\psi_{\rm ini}=|+\cdots+\rangle$ for which $\langle \sigma^z_i(t)\rangle=0$ by symmetry (similar results are obtained with $\psi_{\rm ini}=|\downarrow\cdots\downarrow\rangle$). We focus on the antipodal correlation $\langle \sigma^z_0 \sigma^z_{\lfloor L/2 \rfloor}\rangle$.
The time-evolution is implemented using the $W^{\text{II}}$ algorithm~\cite{zaletel_time-evolving_2015} at order 4, which leads to a Trotter error $\mathcal{O}(\tau^5)$~\cite{bidzhiev_out--equilibrium_2017}, where $\tau$ is the time step. The present calculations have been done with $\tau=0.02$ or $\tau=0.04$ (see the legends). We used the TMS library~\cite{HOUDAYER_TensorMixedStatesJuliaLibrary_2026}, which is built on top of
ITensor~\cite{fishmanITensorSoftwareLibrary2022}. The Ising Hamiltonian is at its critical point $g=1$ and the system has periodic boundary conditions. \cref{fig:mps200} and \cref{fig:mps400} show to what extent such tensor network calculations reproduce the antipodal correlation peak.

Comparing the correlation obtained with MPS with the exact result using the free fermion mapping, we observe that with $\chi=200$ the correlation peak is well reproduced for $L=20$ and $L=30$ but its amplitude and its position in time are not correctly reproduced for $L=40$ and $L=50$. If the maximum bond dimension is increased to $\chi=400$ (\cref{fig:mps400}) the simulation is able to reproduce the peak up to $L=40$ (although some discrepancy with the exact result is visible at the scale of the plot). It should also be remarked that there is hardly any improvement on the correlations for $L=50$ when increasing $\chi$ from 200 to 400, which indicates that much larger values of the bond dimension are probably required to resolve the correlation peak for this system size.

The von Neumann entanglement entropy $S_{\rm vN}(t)$ associated with a central bipartition of the system into two equal halves is plotted in the bottom panels of \cref{fig:mps200} and \cref{fig:mps400}. At short times the entanglement entropy grows linearly with time and is almost independent of the system size. This is the typical behavior in a global quench. Also, the time at which the entropy reaches a maximum is proportional to the system size. When the maximal bond dimension is $\chi=200$ the simulations reproduce quantitatively the exact results up to the time $Jt\simeq 3$ when the entropy reaches $S_{\rm vN}(t)\simeq 4$. Beyond this time the MPS data depart from the exact curves. By doubling the bond dimension to $\chi=400$, we observe a modest improvement, with an entanglement entropy that is correctly reproduced up to $S_{\rm vN}(t)\simeq 4.4$ and $Jt\simeq 3.3$. This captures the entropy maximum for $L=30$ but still fails to capture
the entropy maximum for $L=40$ and $L=50$. With $\chi=600$, the peak height for $L=50$ is still only $\sim50\%$ of its exact value.

From these data we can conclude that simulations with a bond dimension $\chi=400$ and $\chi=600$ would pass the benchmark test up to $L=40$ (with some tolerance $\epsilon \sim 10\%$) but would fail for some system size between 40 and 50.

\section{Scaling of the correlation peak for large systems}
\label[appsec]{app:peak_large_systems}

\begin{figure}
    \centering
    \includegraphics[width=\linewidth]{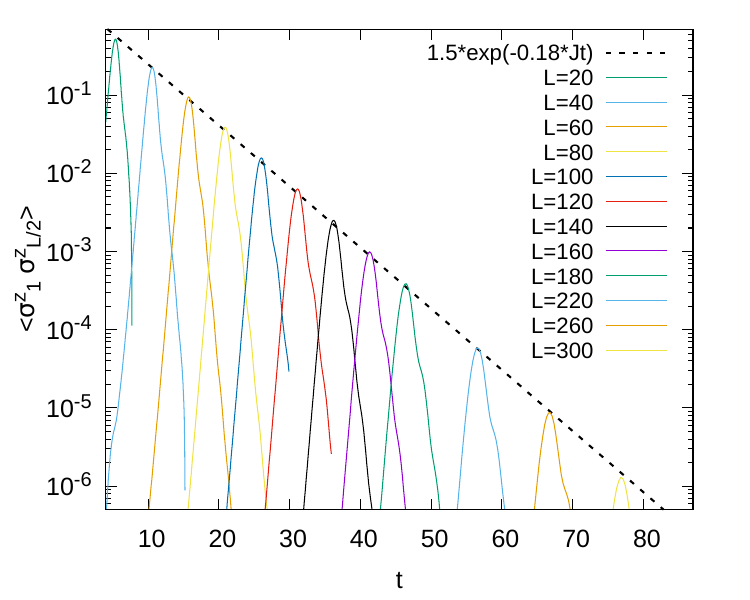}
    \caption{Antipodal correlation peak for large systems up to $L=300$ (exact free fermion results with $|\psi_{\rm ini}\rangle=|+\cdots+\rangle$). The value of correlation between two antipodal spins at the surge time $t_*$ decreases exponentially with $t_*$, and thus with $L$. The dashed line, which corresponds to an exponential decay, is a guide to the eye.
    }
    \label{fig:peak_height_large_systems}
\end{figure}

The free fermion solution allows one to compute spin-spin correlations for large times and large systems. \cref{fig:peak_height_large_systems} represents the antipodal correlation $\langle \sigma^z_0 \sigma^z_{\lfloor L/2 \rfloor}\rangle$ as a function of time and in the vicinity of the surge time, for several system sizes up to $L=300$ and for the initial state $|\psi_{\rm ini}\rangle=|+\cdots+\rangle$. We observe that the height of the peak decreases exponentially with time, and thus with $L$ (since the time $t_*$ of the maximum scales with $L$). A very similar exponential decay is observed when the initial state is $\ket{\psi_{\text{ini}}} = \ket{\downarrow \cdots \downarrow}$.

\printbibliography[heading=bibintoc]

\end{document}